\documentclass[prd,aps,epsf,floats,onecolumn]{revtex4}%
\usepackage{amssymb}
\usepackage{amsfonts}
\usepackage{amsmath}
\usepackage{graphicx}%
\usepackage{algorithm}
\usepackage{algorithmic}
\numberwithin{equation}{section}
\begin{document}
\title{GRADIENT PROFILE ESTIMATION USING EXPONENTIAL CUBIC SPLINE SMOOTHING
IN A BAYESIAN FRAMEWORK}
%\author[1]{Kushani De Silva \corref{cor1}}
%\ead{kdesilva@iastate.edu}
%\author[2]{Carlo Cafaro}
%\author[3]{Adom Giffin}
%\address[1]{Department of Mathematics, Iowa State University, Ames, IA 50011-2104, USA}
%\address[2]{SUNY Polytechnic Institute, 12203 Albany, New York, USA}
%\address[3]{CoR.io Inc, Ontario, Ottawa, Canada}

\author{Kushani De Silva$^{1}$, Carlo
	Cafaro$^{2}$, Adom Giffin$^{3}$, }
\affiliation{$^{1}$Department of Mathematics, Iowa State University, Ames, IA, USA}
\affiliation{$^{2}$SUNY Polytechnic Institute, Albany, NY, USA}
\affiliation{$^{3}$CoR.io Inc, Ontario, Ottawa, Canada}
\begin{abstract}
Attaining reliable profile gradients is of utmost relevance for many physical
systems. In most situations, the estimation of gradient can be inaccurate due to
noise. It is common practice to first estimate the underlying system and then
compute the profile gradient by taking the subsequent analytic derivative. The
underlying system is often estimated by fitting or smoothing the data using
other techniques. Taking the subsequent analytic derivative of an estimated function
can be ill-posed. The ill-posedness gets worse as the noise in the system
increases. As a result, the uncertainty generated in the gradient estimate increases. In
this paper, a theoretical framework for a method to estimate the profile
gradient of discrete noisy data is presented. The method is developed
within a Bayesian framework. Comprehensive numerical experiments are
conducted on synthetic data at different levels of random noise. The accuracy
of the proposed method is quantified. Our findings suggest that the proposed gradient
profile estimation method outperforms the state-of-the-art methods.

Key words: Computational Techniques, Inference Methods, Probability Theory
\end{abstract}

%\pacs{Computational Techniques (02.70-c), Inference Methods (02.50.Tt),
%Probability Theory (02.50.Cw).}
\maketitle

\pagebreak

\section{Introduction}

Estimating the derivative of a system from discrete set of data has a vast
number of applications in many fields such as biology, engineering,
and physics. For instance, determining the particle velocity from
the discrete time-position data in particle image velocimetry and particle
tracking velocimetry experiments \cite{von2011digital,feng2011errors}
are important tasks in plasma physics \cite{chilenski2015improved}.
Applications of velocity estimation in motion control systems using discrete
time data have increased with the invention of microprocessors
\cite{brown1992analysis} (and references therein). Moreover, with the
improvement of technology, faster equipment is now available to
measure high-speed discrete data \cite{von2011digital}.

To estimate the derivative of a system using discrete data, several approaches
have been discussed in the literature. Some of them are finite difference
\cite{feng2011errors} and analytic derivative of either a traditional data
smoothing technique
\cite{reinsch1967smoothing,reinsch1971smoothing,hastie1990generalized,wahba1990spline,wand1994kernel,green1993nonparametric,wold1974spline,silverman1985some}
or Bayesian data smoothing method using spline functions
\cite{wahba1978improper,kimeldorf1970correspondence,ansley1985estimation,berry2002bayesian,fischer2000background}%
. Basically, the latter technique estimates the discrete data using a
smooth function before taking the analytic derivative of the spline function.
Most of the time, cubic splines are used as the smooth function to estimate
the discrete data. However, as pointed out in
\cite{dose2005function,fischer2006flexible}, exponential cubic splines are
superior to cubic splines as they are better at capturing in an
accurate manner abrupt changes in data. Furthermore, exponential cubic spline
is used to directly estimate the unknown velocity in
\cite{von2011digital}. In this case, the spline represent the velocity
instead of discrete data as compared to all previous cases. However, it is
claimed that the noise contaminated in the discrete data can produce huge
uncertainties in the gradient estimates \cite{brown1992analysis}. These
uncertainties tend to magnify in the gradient estimates when the
discrete data are measured in short time intervals, e.g. data captured with
high-speed cameras \cite{von2011digital,feng2011errors}.

Exponential cubic splines has two limiting cases: polygonal function
and cubic spline. These limits are handled by the smoothing parameter.
Therefore, when using exponential cubic spline, special attention must be paid
to the smoothing parameter because extreme values of it can sometimes produce
unrealistic results. The Bayesian approach employed in the study of
\cite{von2011digital} used a Jeffrey's prior on the smoothing parameter simply
to scale down the value to avoid producing drastic results. Without the
optimal smoothness, the spline would not produce optimum gradient estimates.
Moreover in that study, the results are stated only for a small level of
random noise without a quantitative study of the noise factor. The noise in
the data plays a major role in this study and a sensitivity analysis of all
possible noise levels is important. Furthermore, the results in positions,
velocity, and acceleration are not quantitatively evaluated for superiority
in Ref. \cite{von2011digital}.

In our study, motivated by the important work presented in
\cite{von2011digital}, we present a detailed investigation on gradient
profile estimation by employing exponential cubic spline smoothing within a
Bayesian setting. The exponential cubic spline and its properties are
thoroughly studied and a better choice of prior on the smoothing parameters is introduced. Moreover, a comprehensive study of sensitivity analysis of
noise on the gradient estimates is performed. Additionally, the results are
evaluated quantitatively for position, velocity, and acceleration estimates.

The rest of the paper is organized as follows. Section II develops the
idea of separating spaces with mathematical background. The Bayesian framework
of the algorithm is demonstrated in Section III. Afterwards, the computational
details are presented in Section IV and Section V. In particular, in these
sections, our findings are compared to those obtained by means of more
traditional methods such as smoothing data with subsequent analytic derivative. Our concluding remarks are given in Section VI. Finally,
some technical details appear in Appendix \ref{appendix I}.

\section{Separating Spaces in Bayesian Context}

To overcome the dilemma of huge uncertainties in the gradient estimates of
noisy data, the gradient is directly approximated in the space where the
gradient resides using the information available in the space where
data lives. Bayesian methods provide the mechanism to map the information
between the two spaces, data and gradient, by introducing a mathematical model
to relate observable information (data) with the unknown quantity (gradient)
\cite{mr1763essay,box2011bayesian,geweke1989bayesian,dempster1968generalization,gamerman2006markov,knill1996perception,newton1994approximate}%
. According to the problem studied in the paper, the data are the position $x(t)$ of an object
measured at discrete times $t$ while the gradient is the
velocity $v(t)$ of that object. Hence, the mapping between the two spaces
is obtained by the two basic relationships between position
and velocity,%
\begin{equation}
x(t)\overset{\text{def}}{=}\int v(t)dt\text{, and }v(t)\overset{\text{def}}%
{=}\frac{d}{dt}x(t)\text{.} \label{basic relationship}%
\end{equation}
Since the gradient is directly approximated in the velocity space, the mathematical model that claims to represent the velocity of the object is placed in the velocity space. Then mapping the information between the
position and the velocity is performed by integrating the mathematical model using the relation depicted in Eq. ~\eqref{basic relationship}.

Throughout the paper, let the space of the gradient (v-space) be
denoted by $\mathbb{V}$ and the space of the data (x-space) be denoted by
$\mathbb{U}$. Let the noisy data be denoted by $\{t_{i}$,
$x_{i}\}_{i=1}^{n}\in\mathbb{U}$ and the mathematical model by $\mathcal{M}%
\in\mathbb{V}$. As highlighted in the Introduction, the exponential cubic spline is capable of capturing
abrupt changes and therefore it is a better model to represent the velocity of
a fast moving object \cite{dose2005function,fischer2006flexible,von2011digital}. Let us denote
the exponential cubic spline by $S_{v}$ in which the subscript $v$ specifies that
the spline, $S$ is placed in $\mathbb{V}$. As any other spline, the exponential cubic spline is defined in piecewise manner by its function values at a set of knot positions. Since, in our problem, the spline represents the gradient/velocity, the function values are the velocity values defined at the knot positions. Let us denote the velocity values at the knot positions $\xi_{v_i}$ as $f_{v_i}$ for $i=1,2,\cdots,E_v$ with $\xi_{v_1}=t_1$ and $\xi_{v_{E_v}}=t_n$.
Therefore, the exponential cubic spline for our variables,
$\{\xi_{v_{i}}$, $f_{v_{i}}\}_{i=1}^{E_{v}}$ defined in the $i^{th}$ interval, can be written as
\cite{rentrop1980algorithm,fischer2006flexible},
\begin{align}
  &S_{v_{i}}\left(  t\text{, }f_{v}\text{, }\lambda_{v}\text{, }\xi_{v}\text{,
}E_{v}\right)  \overset{\text{def}}{=} f_{v_{i}}\left(  1-h_{v}\right)
+f_{v_{i+1}}h_{v}\label{spline_v}\\
&  +\frac{M_{v_{i}}}{\lambda_{v_{i}}^{2}}\left\{  \frac{\sinh\left[
\mu_{v_{i}}\left(  1-h_{v}\right)  \right]  }{\sinh\left(  \mu_{v_{i}}\right)
}-\left(  1-h_{v}\right)  \right\}  +\frac{M_{v_{i+1}}}{\lambda_{v_{i}}^{2}%
}\left[  \frac{\sinh\left(  \mu_{v_{i}}h_{v}\right)  }{\sinh\left(  \mu
_{v_{i}}\right)  }-h_{v}\right] \nonumber
\end{align}
with $i=1$,..., $E_{v}-1$. The quantities $h_{v}$,
$\mu_{v_{i}}$, and $M_{v_{i}}$ in Eq. (\ref{spline_v}) are
defined as,
\begin{align}
h_{v}\overset{\text{def}}{=}\frac{t-\xi_{v_{i}}}{\xi_{v_{i+1}}-\xi_{v_{i}}%
}\text{, }\mu_{v_{i}}\overset{\text{def}}{=}\lambda_{v_{i}}\left(
\xi_{v_{i+1}}-\xi_{v_{i}}\right)  \text{, and }\\
M_{v_{i}}\overset{\text{def}%
}{=}\left\{  \frac{d^{2}}{dt^{2}}\left[  S_{v_{i}}\left(  t\text{, }%
f_{v}\text{, }\lambda_{v}\text{, }\xi_{v}\text{, }E_{v}\right)  \right]
\right\}  _{t=\xi_{v_{i}}}\text{,} \notag
\end{align}
respectively. Furthermore, $\xi_{v}\overset{\text{def}}{=}\left(
\xi_{v_{1}}\text{,}\cdots\text{, }\xi_{v_{E_{v}}}\right)  $ are the positions of the knots, $f_{v}\overset{\text{def}}{=}\left(  f_{v_{1}}%
\text{,}\cdots\text{, }f_{v_{E_{v}}}\right)  $ are the velocity
values at knots, $\lambda_{v}\overset{\text{def}}{=}\left(  \lambda_{v_{1}}%
\text{,}\cdots\text{, }\lambda_{v_{E_{v}-1}}\right)$ are the tension values between two knots, and, finally, $E_{v}$ are the number of knots of the
spline. Throughout the paper, the subscript $v$ of the variables
specifies that they are placed in $\mathbb{V}$. Using the arrow notations for vectorial quantities, the
exponential cubic spline in Eq.~\eqref{spline_v} can be also viewed in matrix
form as \cite{von2011digital},
\begin{equation}
S_v\left(  \vec{t}\right)  =W\left(  \vec{t}\text{, }\vec{\lambda_{v}}%
\text{,}\vec{\text{ }\xi_{v}}\right)  \vec{f_{v}}\text{.} \label{pappa}%
\end{equation}
In Eq. (\ref{pappa}), $W$ denotes the design matrix
of $\vec{t}$-locations of the data points $\vec{t}$, the $\vec{t}$-locations of the support points $\vec{\xi}_{v}$,
the vector of tension parameters $\vec{\lambda}_{v}$ and, finally,
the vector of function values $\vec{f_{v}}$. Moreover,
the quantity $S_v$ \ is the solution to the variational problem
\cite{von2011digital},
\begin{equation}
\int_{t_{1}}^{t_{n}}\left\{  \left[  \frac{d^{2}S_v\left(  t\right)  }{dt^{2}%
}\right]  ^{2}+\lambda\left(  t\right)  ^{2}\left[  \frac{dS_v\left(  t\right)
}{dt}\right]  ^{2}\right\}  dt\quad\mbox{for}\quad\xi_{i}\leq t\leq\xi_{i+1}.
\end{equation}
The second derivative of $S_{v}$, i.e. $M_{v}$, can be
explicitly found by solving the tri-diagonal system of equations in Eq.
(\ref{pappa}). The function values $\vec{f_{v}}$, the tension values $\vec{\lambda_{v}}$, the positions of the knots $\vec{\text{ }\xi_{v}}
$ and the number of knots $ E_{v}$ are unknown. Therefore, a summary of data and parameters of the model being investigated
is given in Table \ref{data para table}.
\begin{table}[ht]
		\caption{The definition of data and parameters of the system}
	\label{data para table}
	\begin{tabular}{l|l}
		data       & \begin{tabular}[c]{@{}l@{}}$\vec{t}$ and $\vec{x}$ \end{tabular}                                                                   \\ \hline
		parameters & \begin{tabular}[c]{@{}l@{}}$\vec{f}_{v}$, $\vec{\lambda}_{v}$, $\vec{\xi}_{v}$ and $E_v\in\mathbb{Z^{+}}$\end{tabular}
	\end{tabular}
\end{table}

The data resides in $\mathbb{U}$ and the
parameters of the model resides in $\mathbb{V}$ is matched using the relation given in
Eq.~\eqref{basic relationship} with the purpose of establishing
the connection between the two spaces $\mathbb{U}$ and $\mathbb{V}$. In fact,
it matches \textit{data points} in $\mathbb{U}$ with the \textit{spline
values} in $\mathbb{V}$ integrated over time. The relation in
Eq.~\eqref{basic relationship} can be rewritten in terms of our model without loss of generality for an arbitrary $i^{th}$ position as,
\begin{equation}
x_{i}=\int_{t_{1}}^{t_{i}}S_{v}\left(  t\text{, }f_{v}\text{, }\lambda
_{v}\text{, }\xi_{v}\text{, }E_{v}\right)  dt\text{,} \label{pappa3}%
\end{equation}
with \label{spline integration 2} $S_{v%
}\left(  t\text{, }f_{v}\text{, }\lambda_{v}\text{, }\xi_{v}\text{, }%
E_{v}\right)  $ given in Eq.~\eqref{spline_v}. The quantity
$x_{i}$ in Eq. (\ref{pappa3}) can be further reduced to,
\begin{equation}
x_{i}=x_{i-1}+\int_{t_{i-1}}^{t_{i}}   S_{v_{i-1}}\left(  t\text{, }f_{v}\text{,
}\lambda_{v}\text{, }\xi_{v}\text{, }E_{v}\right)  dt\text{,} \label{pappa2}%
\end{equation}
with $i=1$,$\cdots$, $n$. We point out that the exponential cubic
spline is analytically integrable and some technical details are given in
Appendix A. 

\begin{figure}[ptb]
\centering
\begin{minipage}{.5\textwidth}
		\centering
		\makebox{\includegraphics[width=1.0\linewidth]{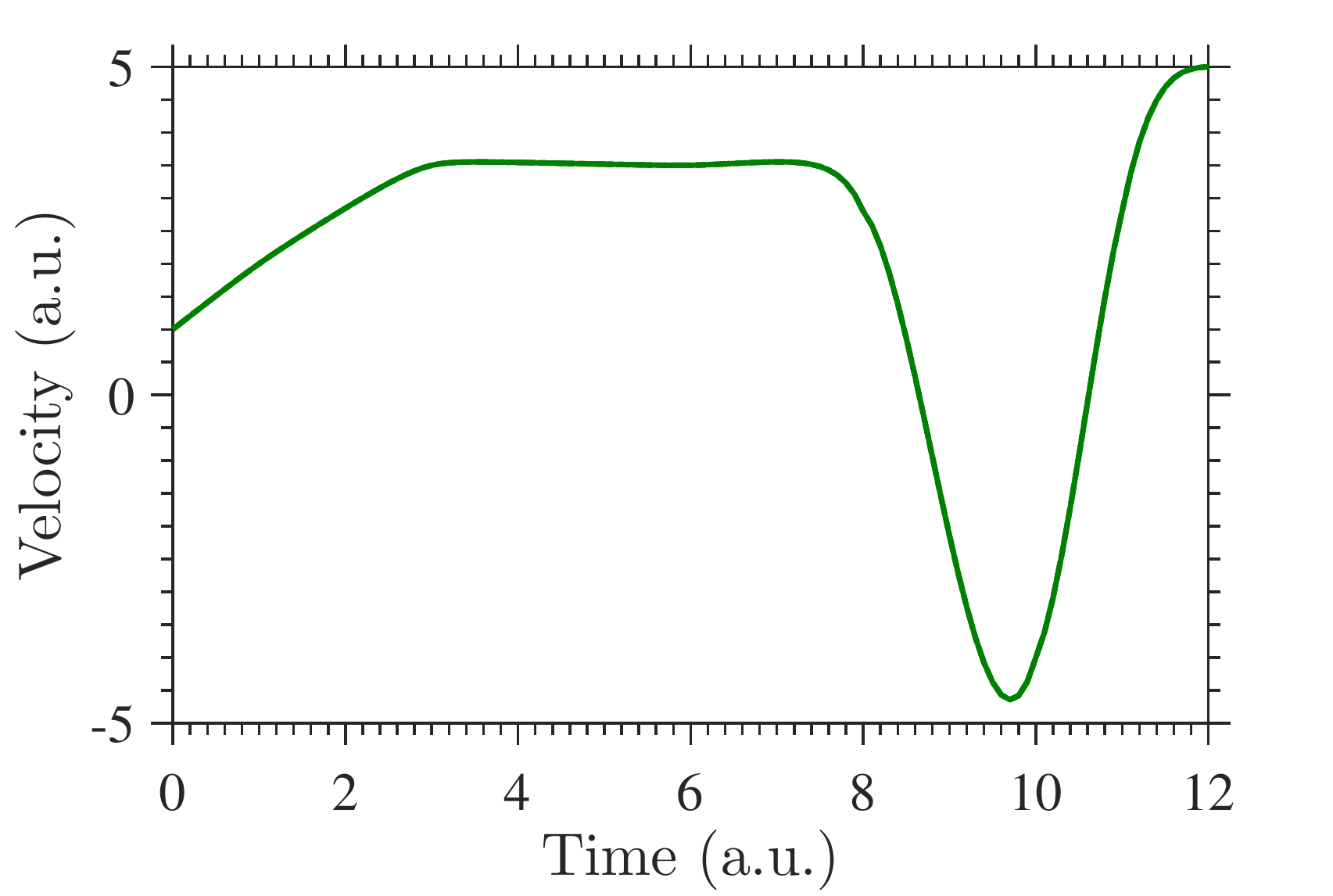}}
		%			\captionof{figure}{A figure}
		\label{fig: timeVelocityPlot}
	\end{minipage}%
 \begin{minipage}{.5\textwidth}
		\centering
		\makebox{\includegraphics[width=1.0\linewidth]{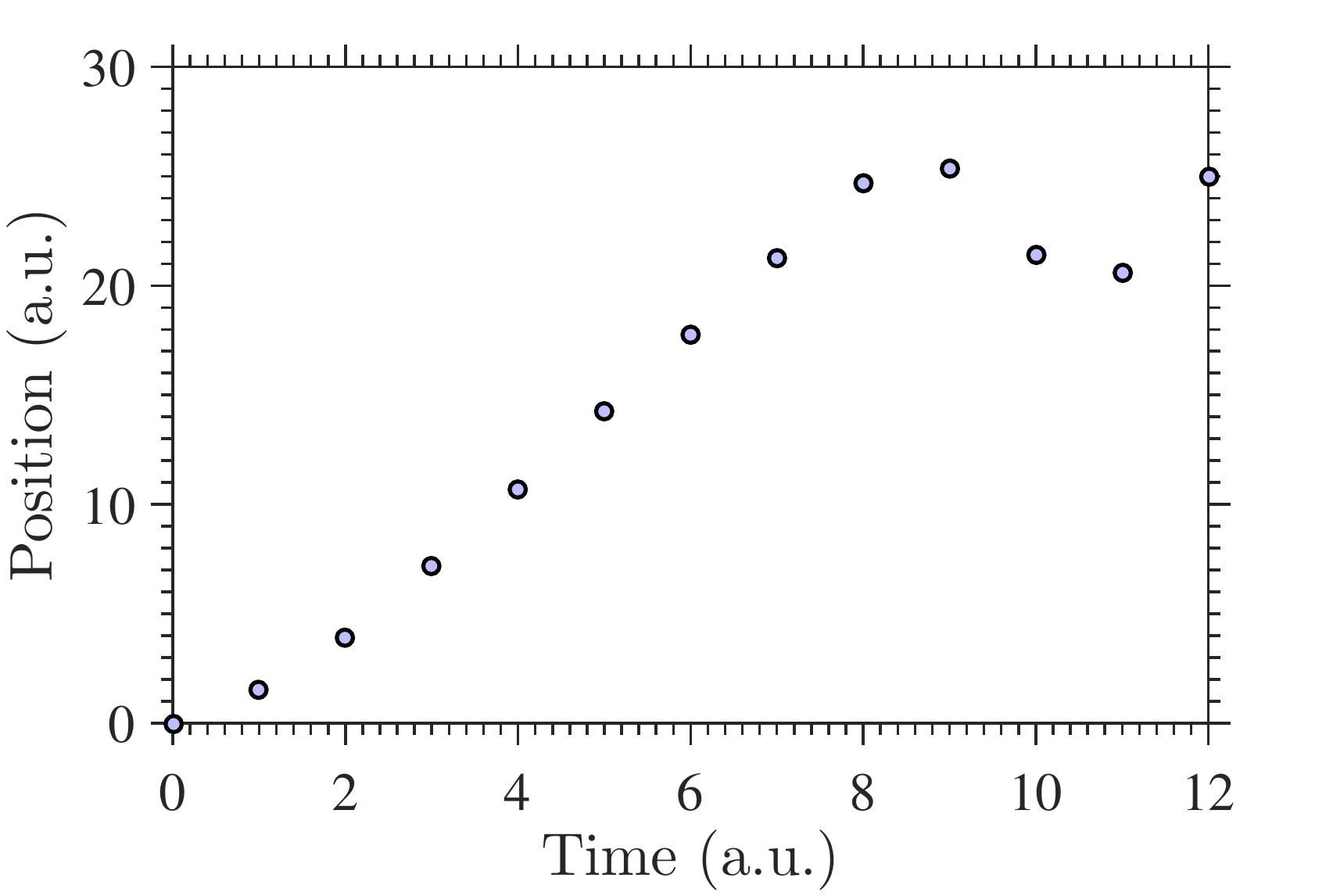}}
		%					\captionof{figure}{Another figure}
		\label{timePositionScatter}
	\end{minipage}
\caption{An example of solving a forward problem from a velocity curve (left) to calculate
positions/distances (right). The velocity curve is the
exponential cubic spline computed at $\vec{\xi}=[0,1,3,6,8,10,11,12]$. The positions were calculated from the velocity using the
forward problem at $n = 13$ different time instances. The time axis is given at arbitrary units (a.u.).}%
\label{time_velocity_position_forwardProblem}%
\end{figure}

When a mathematical model is used to match the unknown quantity and
the observable data, the model should be able to generate data as likely as
possible to the observable information if the desired/unknown
quantity is known. This is called the \emph{forward problem} in Bayesian
language. However, what is required here is to be able to solve the
\emph{inverse problem}. That is, making inferences about the desired/unknown
quantity using the observed information and the model
\cite{mohammad2001bayesian,sivia2006data,gelman2014bayesian}. An example of
solving a forward problem by computing positions when the velocity is known is
shown in Fig. \ref{time_velocity_position_forwardProblem}.

\section{The Bayesian Recipe}

The inverse problem for the position-velocity scenario introduced in the previous section can be solved
by means of Bayes' rule,
\begin{equation}
p\left(  \text{velocity}\in\mathbb{V}|\text{positions}\in\mathbb{U}\right)
=\frac{p(\text{velocity}\in\mathbb{V}) p(\text{positions}\in
\mathbb{U}|\text{velocity}\in\mathbb{V})}{p(\text{positions}\in\mathbb{U})}
\label{bayes}%
\end{equation}
The relation depicted in Eq. (\ref{bayes}) shows how the joint posterior
probability distribution can be built to infer the velocity when the positions are
given. Bayes' rule in Eq. (\ref{bayes}) can be rewritten using data
and parameters given in Table \ref{data para table} as follows,
\begin{equation}
p\left(  \vec{f_{v}}\text{, }\vec{\lambda_{v}}\text{, }\vec{\xi_{v}}\text{,
}E_{v}|\vec{t}\text{, }\vec{x}\text{, }I\right)  =\frac{p(\vec{f_{v}}\text{,
}\vec{\lambda_{v}}\text{, }\vec{\xi_{v}}\text{, }E_{v}|I) p(\vec{x}%
|\vec{f_{v}}\text{, }\vec{\lambda_{v}}\text{, }\vec{\xi_{v}}\text{, }%
E_{v}\text{, }\vec{t}\text{, }I)}{p(\vec{x}|I)}\text{,} \label{bayes2}%
\end{equation}
where the quantity $I$ represents all the relevant
information that concerns the physical scenario under investigation together
with the available knowledge about the experiment being performed. By
assuming the independence in spline variables, the following relation
is satisfied,
\begin{equation}
p(\vec{f_{v}},\vec{\lambda_{v}},\vec{\xi_{v}},E_{v}|I)=p\left(  \vec{f_{v}%
}|I\right)  p\left(  \vec{\lambda_{v}}|I\right)  p\left(  \vec{\xi_{v}%
}|I\right)  p\left(  E_{v}|I\right)  \text{.} \label{ind}%
\end{equation}
Substituting Eq. (\ref{ind}) into Eq. (\ref{bayes2}), we get
\begin{align}
p&\left(  \vec{f_{v}}\text{, }\vec{\lambda_{v}}\text{, }\vec{\xi_{v}}\text{,
}E_{v}|\vec{t}\text{, }\vec{x}\text{, }I\right)  \\
=&\frac{p\left(  \vec{f_{v}%
}|I\right)  p\left(  \vec{\lambda_{v}}|I\right)  p\left(  \vec{\xi_{v}%
}|I\right)  p\left(  E_{v}|I\right)   p(\vec{x}|\vec{f_{v}}\text{, }%
\vec{\lambda_{v}}\text{, }\vec{\xi_{v}}\text{, }E_{v}\text{, }\vec{t}\text{,
}I)}{p(\vec{x}|I)}\text{.} \label{full posterior inverse}\notag
\end{align}
The evidence $p\left(  \vec{x}|I\right)  $ in\ Eq.
(\ref{bayes}) and (\ref{bayes2}) is given by,
\begin{align}
&p\left(  \vec{x}|I\right)  \overset{\text{def}}{=}\\
&\iiiint p\left(  \vec{f_{v}%
}|I\right)  p\left(  \vec{\lambda_{v}}|I\right)  p\left(  \vec{\xi_{v}%
}|I\right)  p\left(  E_{v}|I\right)  p(\vec{x}|\vec{f_{v}}\text{, }%
\vec{\lambda_{v}}\text{, }\vec{\xi_{v}}\text{, }E_{v}\text{, }\vec{t}\text{,
}I)d\vec{f_{v}}d\vec{\lambda_{v}}d\vec{\xi_{v}}dE_{v}\text{,}\notag
\end{align}
where the likelihood distribution is given by $p(\vec{x}|\vec{f_{v}}$,
$\vec{\lambda_{v}}$, $\vec{\xi_{v}}$, $E_{v}$, $\vec{t}$, $I)$. The likelihood
distribution affirms the relation between the data and the spline values.
Since the position data $x_{i}$ with $i=1$,..., $n$
is contaminated with noise, omitting the arrow notation for vectorial quantities,
Eq. (\ref{pappa2}) can be rewritten including the noise term as,
\begin{align}
x_{i}=%
\begin{cases}
x_{1}+\sum_{\gamma=1}^{i-1}\int_{\xi_{v_{\gamma}}}^{\xi_{v_{\gamma+1}}%
}S_{v_{\gamma}}\left(  t\text{, }f_{v}\text{, }\lambda_{v}\text{, }\xi_{v}\text{,
}E_{v}\right)  dt+\int_{\xi_{v_{i}}}^{t_{i}}S_{v_{i}}\left(  t\text{, }%
f_{v}\text{, }\lambda_{v}\text{, }\xi_{v}\text{, }E_{v}\right)  dt\\
+\epsilon_{i}\text{,} \quad\mbox{for }\xi_{v_{i}}<t_{i}<\xi_{v_{i+1}}\\
x_{1}+\sum_{\gamma=1}^{i-1}\int_{\xi_{v_{\gamma}}}^{\xi_{v_{\gamma}+1}%
}S_{v_{\gamma}}\left(  t\text{, }f_{v}\text{, }\lambda_{v}\text{, }\xi_{v}\text{,
}E_{v}\right)  dt+\epsilon_{i}\text{,} \\
\quad\mbox{for }t_{i}=\xi_{v_{i}}%
\end{cases}
\text{,} \label{noised}%
\end{align}
where the noise vector $\vec{\epsilon}\overset{\text{def}}%
{\mathbf{=}}\left(  \epsilon_{1}\text{,..., }\epsilon_{n}\right)  $ is
assumed to be linear and, in addition, is assumed to be Gaussian-distributed
with
\begin{equation}
\vec{\epsilon}\sim N\left(  \vec{\mu},\Sigma\right)  \text{.}
\label{gauss noise}%
\end{equation}
In Eq. (\ref{gauss noise}), the quantity $\vec{\mu}$
is the $\left(  n\times1\right)  $-dimensional mean vector
while $\Sigma$ is a $\left(  n\times n\right)  $-dimensional
covariance matrix of the noise. Furthermore, the noise is assumed to be uncorrelated. In
other words, $\Sigma$ is a diagonal matrix with diagonal elements
given by $\left\{  \sigma_{1}\text{,}\cdots\text{, }\sigma_{n}\right\}  $. It
is also assumed that $\vec{\mu}=\vec{0}$. From Eq.
(\ref{noised}), the following equation can be written for the $i$-th
noise term,
\begin{equation}\label{likelihood-noise}
\epsilon_{i}=x_{i}-\left(  x_{1}+\sum_{\gamma=1}^{i-1}\int_{\xi_{v_{\gamma}}%
}^{\xi_{v_{\gamma+1}}}S_{v_{\gamma}}\left(  t\text{, }f_{v}\text{, }\lambda
_{v}\text{, }\xi_{v}\text{, }E_{v}\right)  dt+\int_{\xi_{v_{i}}}^{t_{i}%
}S_{v_{i}}\left(  t\text{, }f_{v}\text{, }\lambda_{v}\text{, }\xi_{v}\text{,
}E_{v}\right)  dt\right)  \text{,}\quad
\end{equation}
where $i=1$,$\cdots$, $n$. The symbols $S_{v_{\gamma}}$ and $S_{v_i}$ denote the exponential spline function in Eq. (\ref{spline_v}) in the $\gamma^{th}$ and the $i^{th}$ sub-intervals, respectively.

 Assuming that for all $i=1$,..., %
$n$ the data values are independent and identically distributed, the
likelihood distribution can be rewritten as
\begin{equation}
p(\vec{x}|\vec{f_{v}}\text{, }\vec{\lambda_{v}}\text{, }\vec{\xi_{v}}\text{,
}E_{v}\text{, }\vec{t}\text{, }I)=\prod_{i=1}^{n}p(x_{i}|\vec{f_{v}}\text{,
}\vec{\lambda_{v}}\text{, }\vec{\xi_{v}}\text{, }E_{v}\text{, }t_{i}\text{,
}I)\text{.} \label{likelihood-product}%
\end{equation}
Combining Eqs.~\eqref{likelihood-noise} and~\eqref{likelihood-product}
together with the additional assumption that $\sigma_{i}=\sigma_{e}%
=$constant for any $1\leq i\leq n$, the likelihood
distribution becomes (case I of Eq.~\eqref{noised} is assumed here for generality),
\begin{align}
&  p(\vec{x}|\vec{f_{v}}\text{, }\vec{\lambda_{v}}\text{, }\vec{\xi_{v}%
}\text{, }E_{v}\text{, }\vec{t}\text{, }I)   =\prod_{i=1}^{n}\frac{1}{\sqrt
{2\pi\sigma_e^{2}}}\exp\left[  -\frac{1}{2\sigma_{e}^{2}}\left(  \epsilon
_{i}\right)  ^{2}\right] \nonumber\\
&  =\left(  2\pi\sigma_{e}^{2}\right)  ^{-n/2}      \exp\sum_{i=1}^{n}\left\{
-\frac{1}{2\sigma_{e}^{2}}\left[  x_{i}-\left(  x_{1}+\sum_{\gamma=1}%
^{i-1}\int_{\xi_{v_{\gamma}}}^{\xi_{v_{\gamma+1}}}S_{v_{\gamma}}(\cdot)dt+\int_{\xi
_{v_{i}}}^{t_{i}}S_{v_{i}}(\cdot)dt\right)  \right]  ^{2}\right\} \nonumber\\
&  =\left(  2\pi\sigma_{e}^{2}\right)  ^{-n/2}\exp\sum_{i=1}^{n}%
\mathcal{G}(x_{i}\text{,}\vec{\text{ }f_{v}}\text{,}\vec{\text{ }\lambda_{v}%
}\text{, }\vec{\xi_{v}}\text{, }E_{v}\text{, }\vec{t})\text{,}
\label{likelihood1}%
\end{align}
with 
\begin{equation}
\mathcal{G}(x_{i}\text{,}\vec{\text{ }f_{v}}\text{,}\vec{\text{ }\lambda_{v}%
}\text{, }\vec{\xi_{v}}\text{, }E_{v}\text{, }\vec{t})\overset{\text{def}}%
{=}-\frac{1}{2\sigma_{e}^{2}}\left[  x_{i}-\left(  x_{1}+\sum_{\gamma=1}%
^{i-1}\int_{\xi_{v_{\gamma}}}^{\xi_{v_{\gamma+1}}}S_{v_{\gamma}}(\cdot)dt+\int_{\xi
_{v_{i}}}^{t_{i}}S_{v_{i}}(\cdot)dt\right)  \right]  ^{2}\text{.} \label{G}%
\end{equation}
When the number of knots is known ahead of time, the product of prior
distributions in Eq. (\ref{ind}) can be rewritten as
\begin{equation}
p(\vec{f_{v}}\text{, }\vec{\lambda_{v}}\text{, }\vec{\xi_{v}}\text{, }%
E_{v}|I)=p\left(  \vec{f_{v}}|E_{v}\text{, }I\right)  p\left(  \vec
{\lambda_{v}}|E_{v}\text{, }I\right)  p\left(  \vec{\xi_{v}}|E_{v}\text{,
}I\right)  \text{.} \label{ind2}%
\end{equation}
%talking about only the Gaussian prior
\subsection{The choice of prior probability distributions}
The shape of the curve produced between knots depends on the values of $\vec{\lambda}_{v}$. One of the concerns with the tension parameter is
that it is a scale parameter and it can assume any real
number. The shape of the curve between two knots can change from cubic function to a polygonal function when its value goes from $0$ to $\infty$ \cite{rentrop1980algorithm,fischer2006flexible}. A polygonal function is not a good representation of a moving object due to its
lack of smoothness. Thus, the value of the tension plays a big role in getting the perfect velocity curve as explained by the observed data. Therefore, prior distributions on $\lambda_v$ should be chosen carefully to control its value.

Previous works in the literature
\cite{fischer2006flexible,fischer2006non} has used Jeffreys' prior to
scale down large values produced for the tension. In this paper, we have employed a Gaussian distribution as a prior for the tension. It allows the
choice of sensible values with a higher probability and non-sensible values
with a lower probability. A bonus of choosing a Gaussian prior is that it is a
conjugate prior of the Gaussian likelihood and, as a consequence,
makes the resulting posterior a member of a family of the Gaussian
distributions. 

Assuming identical and independent tension parameters $\vec
{\lambda_{v}}$, the prior distribution $p(\vec{\lambda_{v}}|E_{v}%
$, $I)$ in Eq. (\ref{ind2}) can be written as
\begin{align}
p(\vec{\lambda_{v}}|E_{v}\text{, }I)  &  =\prod_{i=1}^{E_{v}-1}p\left(
\lambda_{v_{i}}|E_{v}\text{, }I\right) \nonumber\\
&  =\prod_{i=1}^{E_{v}-1}\frac{1}{\sqrt{2\pi\sigma_{\theta_{\lambda_i}}^{2}}}\exp\left[
-\frac{\left(  \lambda_{v_{i}}-\mu_{i}\right)  ^{2}}{2\sigma_{\theta_{\lambda_i}}^{2}}\right]
\nonumber\\
&  =\left(  2\pi\sigma_{\theta_{\lambda}}^{2}\right)  ^{-(E_{v}-1)/2}\exp\left[  -\sum
_{i=1}^{E_{v}-1}\frac{\left(  \lambda_{v_{i}}-\mu_{i}\right)  ^{2}}%
{2\sigma_{\theta_{\lambda}}^{2}}\right] \\
& (\text{by taking  }\sigma_{\theta_{\lambda_i}}=\sigma_{\theta_{\lambda}}\forall
i \text{ with } 1\leq i\leq E_{v}-1)\text{,}\notag  \label{lambda Gaussian prior}
\end{align}
with $\mu_i$ and $\sigma_{\theta_{\lambda}}$ representing the mean and the standard deviation of the parameter $\lambda_{v_i}$.
The prior distributions $p\left(  \vec{f_{v}}|E_{v}\text{, }I\right)  $ and
$p\left(  \vec{\xi_{v}}|E_{v}\text{, }I\right)  $ in Eq. (\ref{ind2})
are assumed to be constant as no prior information on the behavior was identified. Finally, using Eqs.
(\ref{likelihood1}) and (\ref{lambda Gaussian prior}), the posterior
distribution $p\left(  \vec{f_{v}}\text{,}\vec{\text{ }\lambda_{v}}\text{,
}\vec{\xi}\text{, }E_{v}|\vec{t}\text{, }\vec{x}\text{, }I\right)  $
in\ Eq. (\ref{bayes2}) can be viewed as,
\begin{align}
p\left(  \vec{f_{v}}\text{,}\vec{\text{ }\lambda_{v}}\text{, }\vec{\xi}\text{,
}E_{v}|\vec{t}\text{, }\vec{x}\text{, }I\right)   &  \propto p\left(
\vec{\lambda_{v}}|E_{v}\text{, }I\right)  p(\vec{x}|\vec{f_{v}}\text{, }%
\vec{\lambda_{v}}\text{, }\vec{\xi}\text{, }E_{v}\text{, }\vec{t}\text{,
}I)\nonumber\\
&  \propto\left(  2\pi\sigma_{\theta_{\lambda}}^{2}\right)  ^{-(E_{v}-1)/2+(-n/2)}\\
&\exp\left[
-\sum_{i=1}^{E_{v}-1}\frac{\left(  \lambda_{v_{i}}-\mu_{i}\right)  ^{2}%
}{2\sigma_{\theta_{\lambda}}^{2}} + \sum_{i=1}^{n}\mathcal{G}\left(  x_{i}%
\text{,}\vec{\text{ }f_{v}}\text{, }\vec{\lambda_{v}}\text{, }\vec{\xi}\text{,
}E_{v}\text{, }\vec{t}\right)\right] \notag
\end{align}
where $\mathcal{G}(x_{i}$,$\vec{\text{ }f_{v}}$,$\vec{\text{ }%
\lambda_{v}}$, $\vec{\xi_{v}}$, $E_{v}$, $\vec{t})$ is given in Eq.
(\ref{G}).

%%%%%%%%%%

\section{Methodology}

Synthetic data of
time-positions were generated from the same velocity curve used in the forward problem shown in Fig.
\ref{time_velocity_position_forwardProblem},
characterized by a reasonably large sample size with %
$n=121$. Then, the Gaussian noise with
$\mu=0$ and $\sigma$ was added to the data,%
\begin{equation}
x(t)=x_{\text{true}}(t)+\left[  \mu+z\sigma\right]  \text{,}
\label{noise creation}%
\end{equation}
with $z$ being the standard parameter introduced when dealing
with normalized Gaussian data sets. Following from the posterior
distributions defined in Section III, the number of parameters of the system
is determined by $E_{v}$. Furthermore, when the positions and
the number of knots are provided by the user, the total number of parameters
of the system will be $2E_{v}-1$.\\

For the synthetic data, following the work in Ref.
\cite{rentrop1980algorithm}, we select $E_{v}=8$ knots at $\vec
{\xi_{v}}=[0,1,3,6,8,10,11,12]$ in to inferring the spline. Then, the
posterior distribution with different priors (see Eq.
~\eqref{full posterior inverse}) were simulated using a hybrid MCMC
algorithm called DRAM \cite{haario2006dram}. The simulation begins
with an initial minimization process of the negative of the posterior
distributions. The initial values of the parameters used were,
\begin{equation}
f_{v_{\text{initial}}}=0\text{, and }\lambda_{v_{\text{initial}}}=0.001.
\end{equation}
We remark that the posterior distribution is undefined at %
$\lambda_{v}=0$ and, in addition, the smallest value that MATLAB
\cite{MATLAB:2015} could handle for $\lambda_{v}$ was experimentally
found to be equal to $0.001$. The initial minimization acts as a
jump start to the DRAM algorithm. The proposal distributions
of the parameters in DRAM that generate the candidate sample values are
Gaussian distributions. The user must input values for the parameters of the
proposal distributions (i.e., mean and variance of Gaussian distributions) as given in Table \ref{table: DRAM proposal}
with $LB$ and $UB$ being the lower and upper bounds of the parameters,
respectively.

\begin{table}[ht]
\caption{\textit{Initial proposal distributions of the system parameters.}}%
\label{table: DRAM proposal}%
\centering
\fbox{
\begin{tabular}
[c]{l|l|l}%
parameter & proposal distribution & \begin{tabular}[c]{@{}c@{}}prior \\ distribution\end{tabular}\\\hline
$f_{v_{i}}$ for $i=1,\cdots,E_{v}$ & Gaussian($\mu_{v_{i}},\sigma_{v_{i}}^{2}%
$) $\in(LB_{v_{i}},UB_{v_{i}})$ & $\mu_{\theta_{v_{i}}},\sigma_{\theta_{v_{i}%
}}$\\
$\lambda_{v_{i}}$ for $i=1,\cdots,E_{v}-1$ & Gaussian($\mu_{\lambda_{i}%
},\sigma_{\lambda_{i}}^{2}$) $\in(LB_{\lambda_{i}},UB_{\lambda_{i}})$ &
$\mu_{\theta_{\lambda_{i}}},\sigma_{\theta_{\lambda_{i}}}$%
\end{tabular}
}\end{table}

The optimum values $\vec{f_{v}}^{\ast}$ and%
$\vec{\text{ }\lambda_{v}}^{\ast}$ determined by the initial
minimization were used as the mean values of the proposal distributions, so that the initial proposal distributions are centered around
$\vec{f_{v}}^{\ast}$ and$\vec{\text{ }\lambda_{v}}^{\ast}$. The information of prior distributions can be specified with prior means $\mu_{\theta_{v_{i}}}$, $\mu_{\theta_{\lambda_{i}}}$ and prior variances $\sigma_{\theta_{v_{i}%
}}$,
$\sigma_{\theta_{\lambda_{i}}}$. The DRAM procedure takes the variance of the prior as
the initial variance of the parameters if the covariance matrix is not
specified by the user. When the prior variance is infinity, the
variances of the proposal distributions are calculated as,
\begin{equation}
\sigma_{v_{i}}^{2}=\left(  \mu_{v_{i}}\times0.05\right)  ^{2}\text{, and
}\sigma_{\lambda_{i}}^{2}=\left(  \mu_{\lambda_{i}}\times0.05\right)
^{2}\text{.}%
\end{equation}
In other words, if no prior information is available about the
parameter, $5\%$ of the mean of the initial distribution is taken as the
initial variance. Finally, samples were drawn from these Gaussian
proposals and the marginal densities of the parameters were computed.
The MCMC simulations were run until convergence was observed. Each
sampling procedure of length $\mathrm{MCMCnsimu}$ was repeated $\mathrm{MCMCN}%
=100$ times. At each of the $\mathrm{MCMCN}$-th repetitions,
the random noise with the same mean and the same standard deviation as reported in
Eq. (\ref{noise creation}) was added. Hence, a different set of data
with the same level of noise was used at each iteration. At each
$\mathrm{MCMCN}$-th repetition, the initial minimization process was
newly ran. All the sample chains were stored and, in addition, the
first few samples were destroyed using a $\mathrm{MCMCnsimu}\ast
p\%$ burn-in period. The quantity $p$ denotes the burn-in time parameter and, in our case is specified by the interval $(0,100)$. The average marginal densities were then
computed for each parameter using the remaining samples. In what
follows, we assume that \textrm{nsimu} denotes the remaining number
of samples after burn-in period, that is, $\mathrm{nsimu}\overset{\text{def}%
}{=}(1-p)\%\ast\mathrm{MCMCnsimu}$. Then, for any $k=1$,$\cdots$, $E_{v}$, the
mean values $\left\{  f_{v_{k}}^{\ast}\right\}  $ of those mean marginal
densities were calculated as,
\begin{equation}
f_{v_{k}}^{\ast}\overset{\text{def}}{=}\frac{1}{\mathrm{nsimu}}\sum
_{i=1}^{\mathrm{nsimu}}\overline{f_{v}}_{ki}\text{,} \label{f1}%
\end{equation}
where$\overline{\text{ }f_{v}}_{ki}$ denote the average
sample values of $f_{v_k}$ after \textrm{MCMCN} number of iterations,%
\begin{equation}
\overline{f_{v}}_{ki}\overset{\text{def}}{=}\frac{1}{\mathrm{MCMCN}}\sum
_{j=1}^{\mathrm{MCMCN}}f_{v_{kij}}\text{.} \label{f2}%
\end{equation}
 Following the same line of reasoning presented for Eqs. (\ref{f1})
and (\ref{f2}), we also have%
\begin{equation}
\lambda_{v_{s}}^{\ast}\overset{\text{def}}{=}\frac{1}{\mathrm{nsimu}}%
\sum_{i=1}^{\mathrm{nsimu}}\overline{\lambda_{v}}_{si}=\frac{1}{\mathrm{nsimu}%
\times\mathrm{MCMCN}}\sum_{i=1}^{\mathrm{nsimu}}\sum_{j=1}^{\mathrm{MCMCN}%
}\lambda_{v_{sij}}\text{,}%
\end{equation}
for any $s=1$,$\cdots$, $E_{v}-1$ with $\overline
{\lambda_{v}}_{si}$ denoting the average sample values of tension after
MCMCN number of iterations,%
\begin{equation}
\overline{\lambda_{v}}_{si}\overset{\text{def}}{=}\frac{1}{\mathrm{MCMCN}}%
\sum_{j=1}^{\mathrm{MCMCN}}\lambda_{v_{sij}}\text{.}%
\end{equation}
Finally, the standard deviations $\left\{  \sigma_{v_{k}}^{\ast
}\right\}  $ and $\left\{  \sigma_{\lambda_{s}}^{\ast}\right\}  $ of
the parameters are calculated from the average of all chains after
\textrm{MCMCN} iterations as,
\begin{equation}
\sigma_{v_{k}}^{\ast}\overset{\text{def}}{=}\sqrt{\frac{1}{\mathrm{nsimu}%
-1}\sum_{i=1}^{\mathrm{nsimu}}\left(  \overline{f_{v}}_{ki}-f_{v_{k}}^{\ast
}\right)  ^{2}}\text{,}%
\end{equation}
for any $k=1$,$\cdots$, $E_{v}$ and,
\begin{equation}
\sigma_{\lambda_{s}}^{\ast}\overset{\text{def}}{=}\sqrt{\frac{1}%
{\mathrm{nsimu}-1}\sum_{i=1}^{\mathrm{nsimu}}\left(  \overline{\lambda
}_{v_{si}}-\lambda_{v_{s}}^{\ast}\right)  ^{2}}\text{,}%
\end{equation}
for any $s=1$,$\cdots$, $E_{v}-1$, respectively.

\section{Results}

This section presents the results obtained from the method explained in the previous sections. The sensitivity of gradient estimates on the noise levels of the position data are tested. The noise levels on x-space is depicted in Fig. \ref{timePositionScatter} with different noise
levels are denoted by $\sigma_{e}$. The noise of same level was added randomly at the beginning of the optimization process with keeping all other variables same:
The data size $n$, the number of knots of the exponential cubic
spline $E_{v}$, the positions of the knots $\vec{\xi_{v}}$,
the number of MCMC iterations $\mathrm{MCMCN}$, and the true data
positions. 
The added noise have five different levels
of $\sigma_{e}\in\left\{  0.001\text{, }0.3\text{, }0.7\text{, }1.0\text{,
}1.3\right\}$ from a standard Gaussian distribution. By the
definition of the Gaussian distribution, the first four noise levels have a
probability of $68.2\%$ while the last noise level has a
probability of $95\%$.The noise levels are selected so that
that they represent distinct possibilities of the Gaussian noise.

The sensitivity of noise levels was investigated for the posterior
distribution with the Gaussian prior used for the tension
parameter. These results of the sensitivity analysis were compared
against a common traditional method of fitting the same type of exponential cubic
spline in $\mathbb{U}$ and the gradient is obtained by analytically differentiating the
resulting fitted spline function. The notation \textquotedblleft$i$-fit ($i$-space)
with $i=$ x, v\textquotedblright\ denotes the fact that the
$i$-fit was fitted on the $i$-space where the x-fit is the distance
profile while the v-fit is the gradient profile.

\begin{figure}[ptb]
\centering
\begin{minipage}{.5\textwidth}
		\centering
		\makebox{\includegraphics[width=1.0\linewidth]{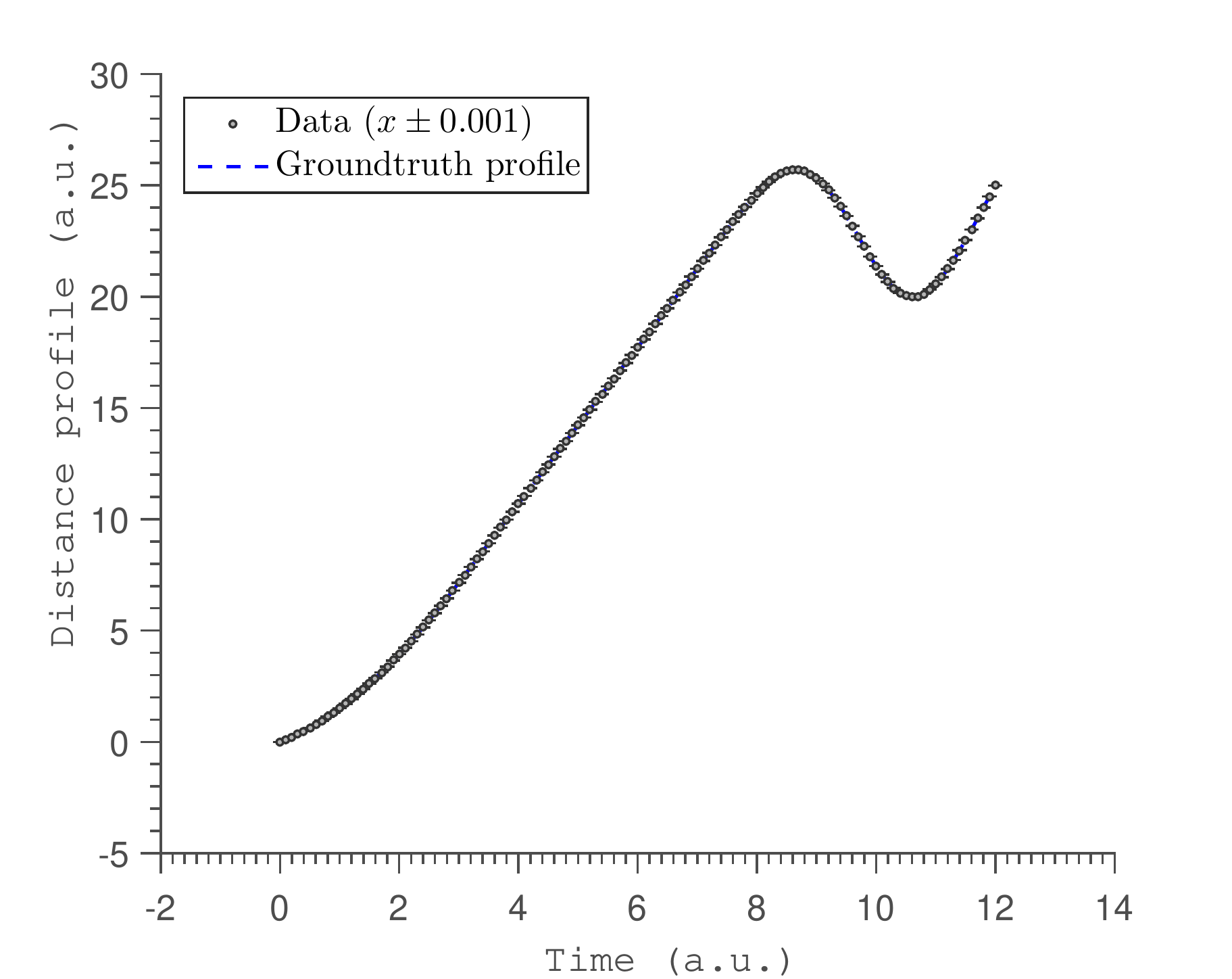}}
		%			\captionof{figure}{A figure}
		\label{fig:sd_0_001_data_xspace}
	\end{minipage}%
 \begin{minipage}{.5\textwidth}
		\centering
		\makebox{	\includegraphics[width=1.0\linewidth]{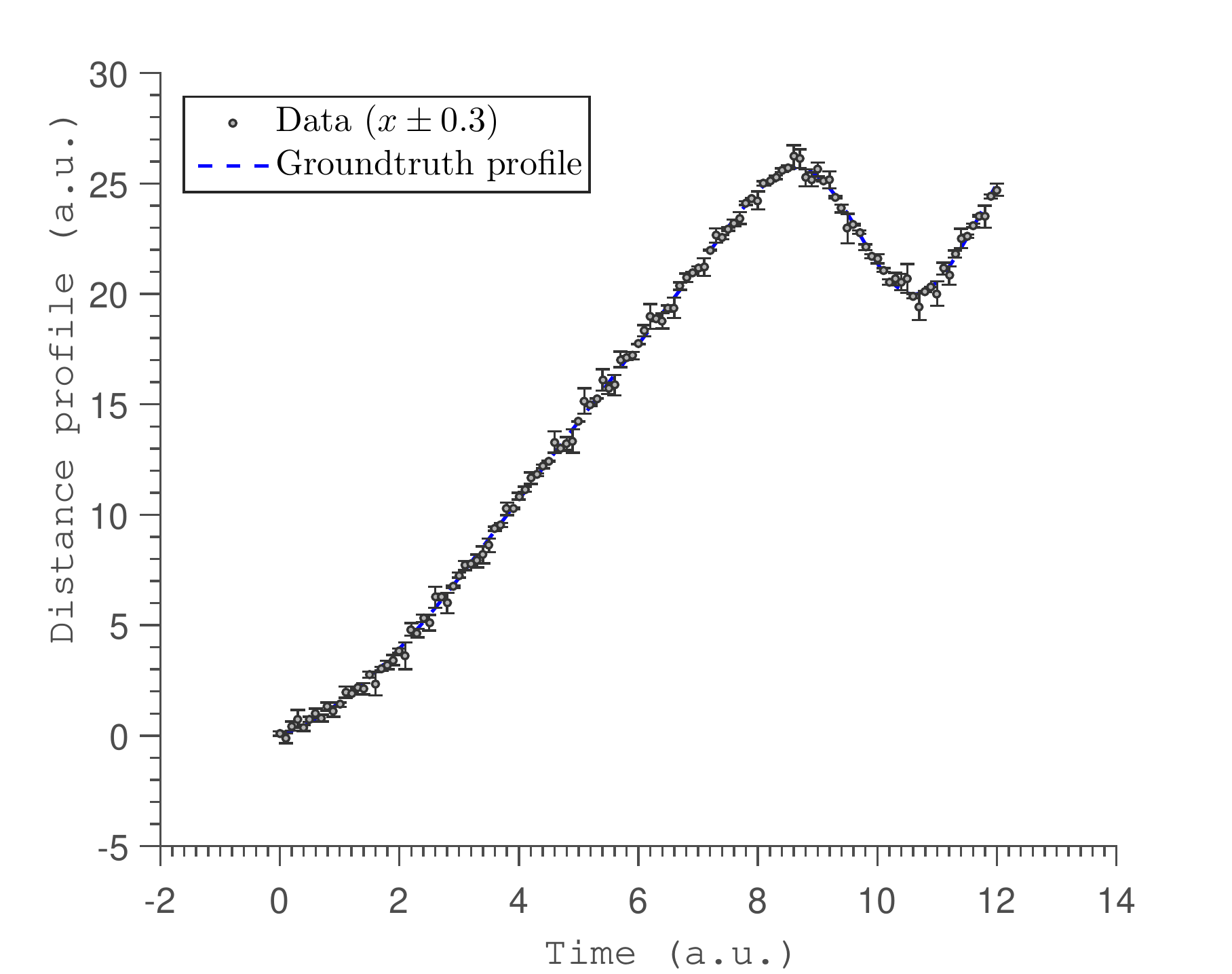}}
		%					\captionof{figure}{Another figure}
		%		\label{fig:sd_0_3_data_xspace}
	\end{minipage}
\begin{minipage}{.5\textwidth}
		\centering
		\makebox{\includegraphics[width=1.0\linewidth]{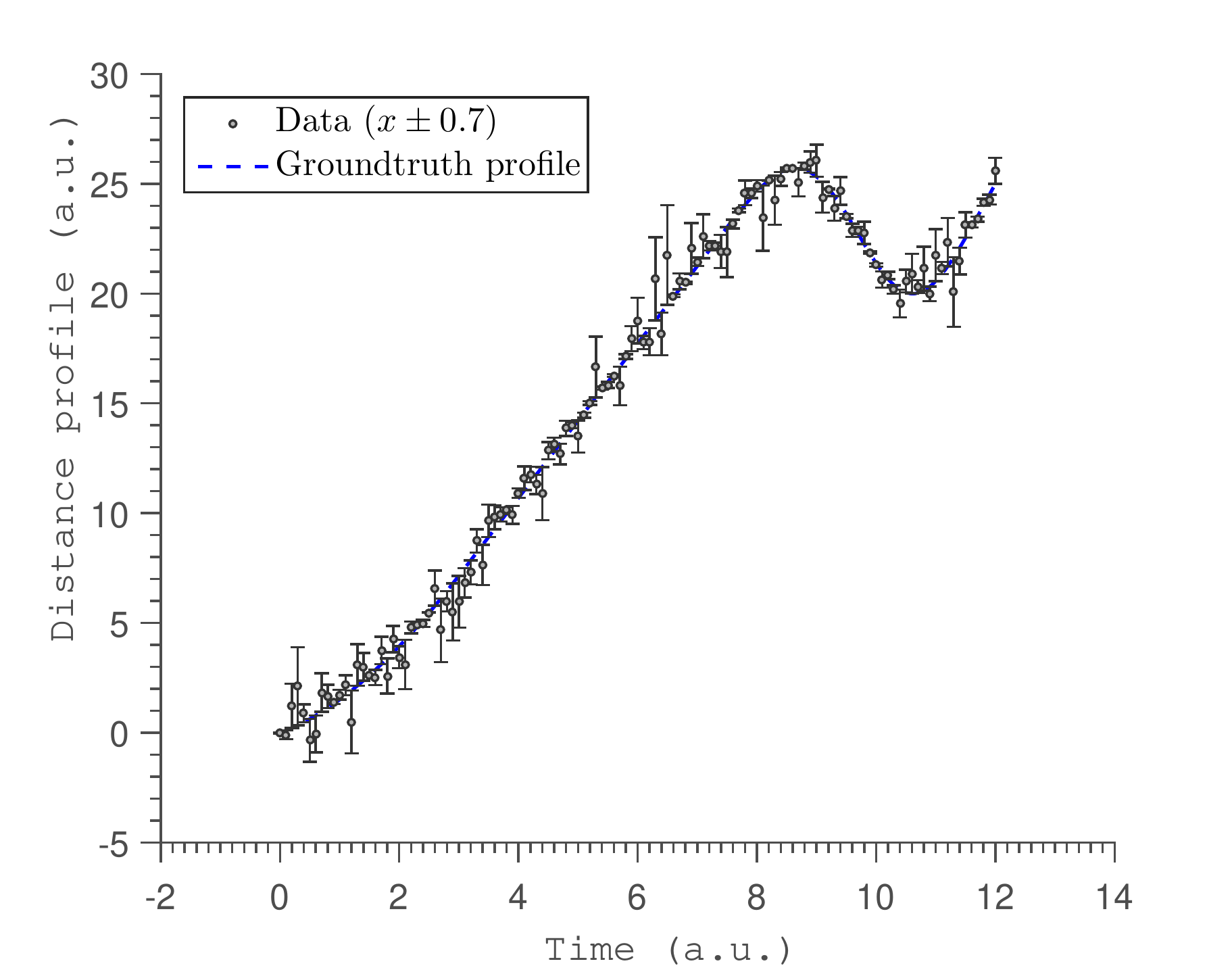}}
		%		\captionof{figure}{A figure}
		%		\label{fig:sd_0_7_data_xspace}
	\end{minipage}%
 \begin{minipage}{.5\textwidth}
		\centering
		\makebox{\includegraphics[width=1.0\linewidth]{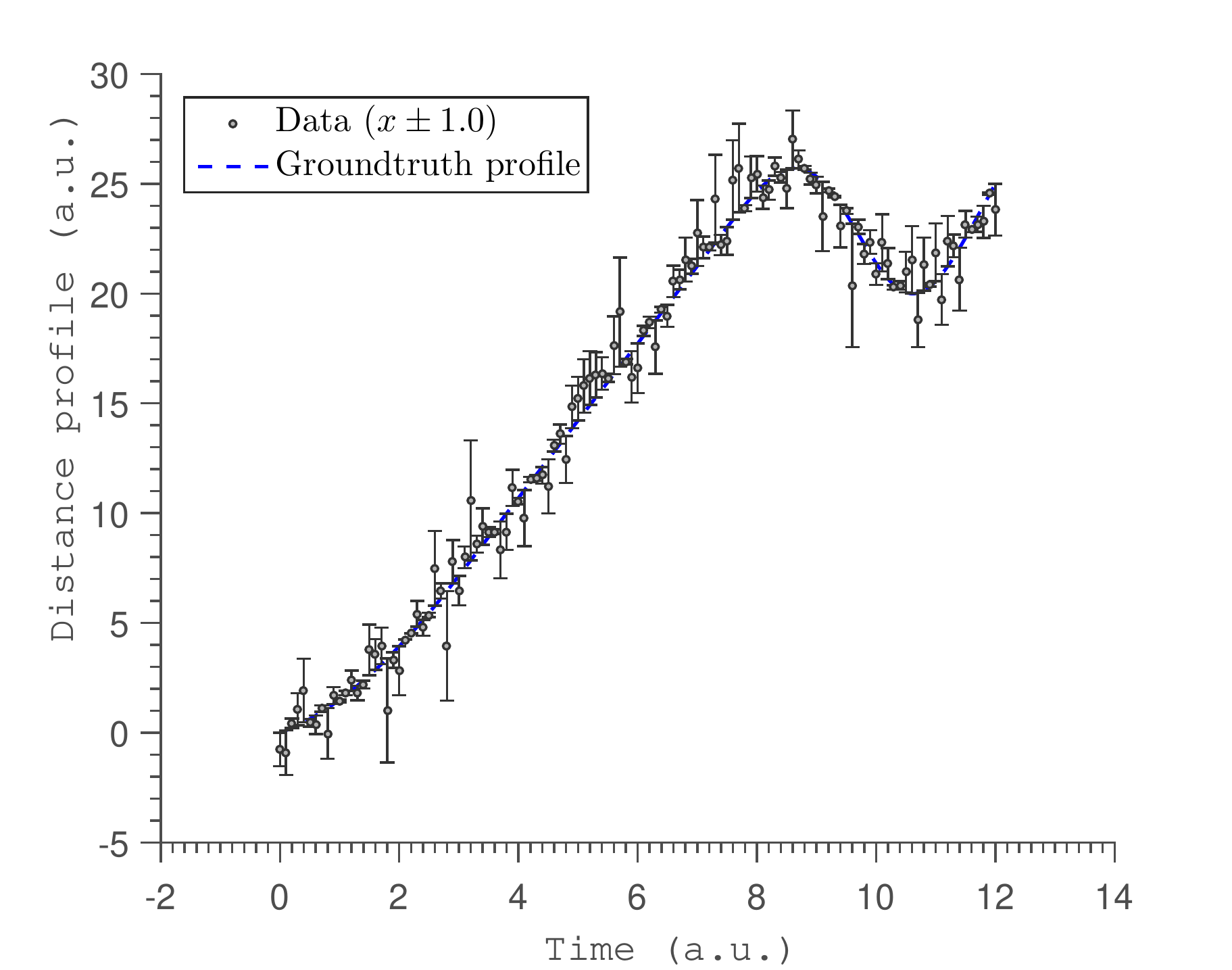}}
		%		\captionof{figure}{Another figure}
		%		\label{fig:sd_1_0_data_xspace}
	\end{minipage}
\begin{minipage}{.5\textwidth}
		\centering
		\makebox{\includegraphics[width=1.0\linewidth]{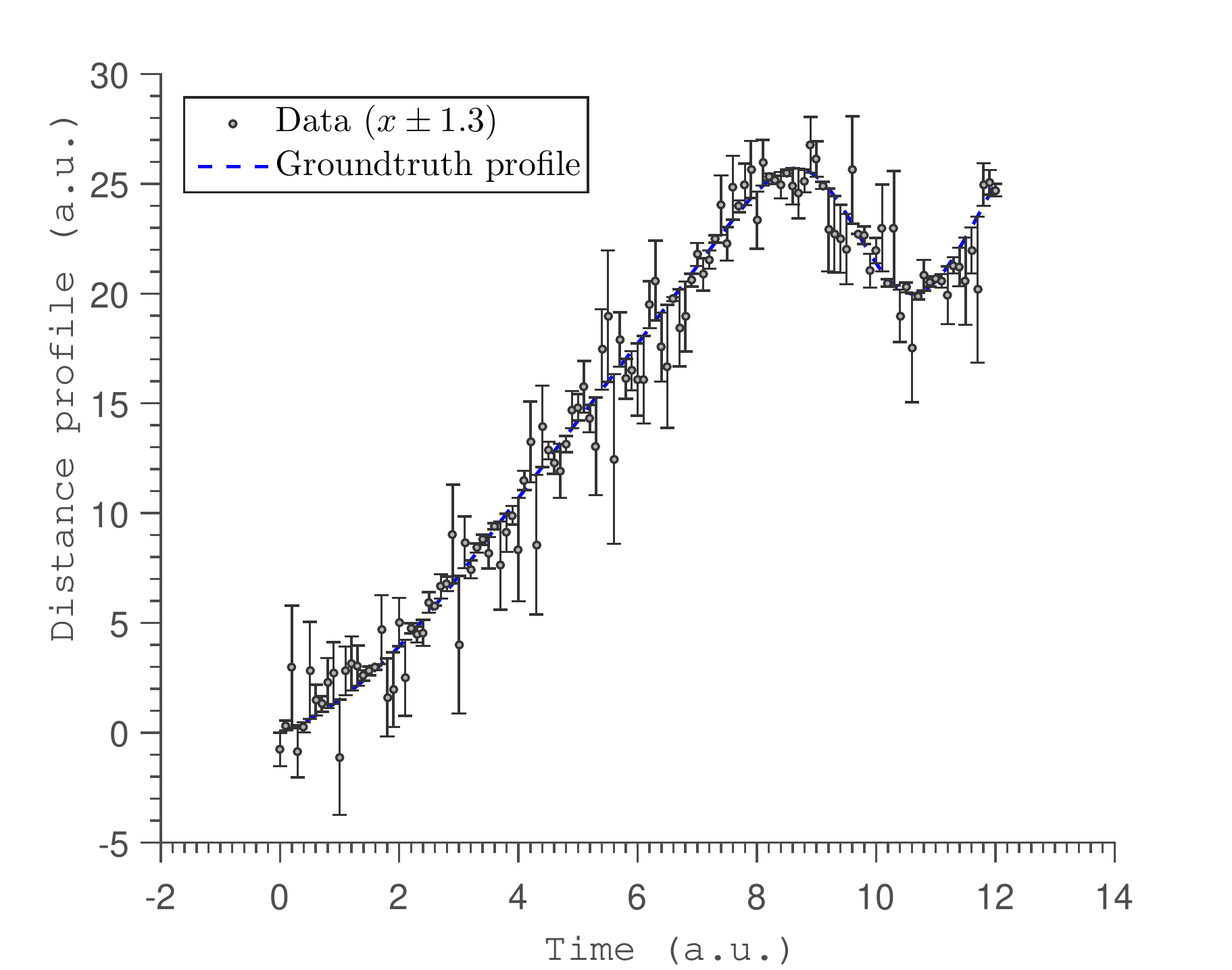}}
	\end{minipage}
\caption{The panel of time-position data with error bars at five different
noise levels. Noise levels increase from left to right and top to bottom. The
length of error bars is relative to the magnitude of the noise.}%
\label{allnoise_data_xspace}%
\end{figure}
\begin{figure}[ptb]
\centering
%	\captionsetup{justification=centering}
\makebox{\includegraphics[width=0.3\textwidth]{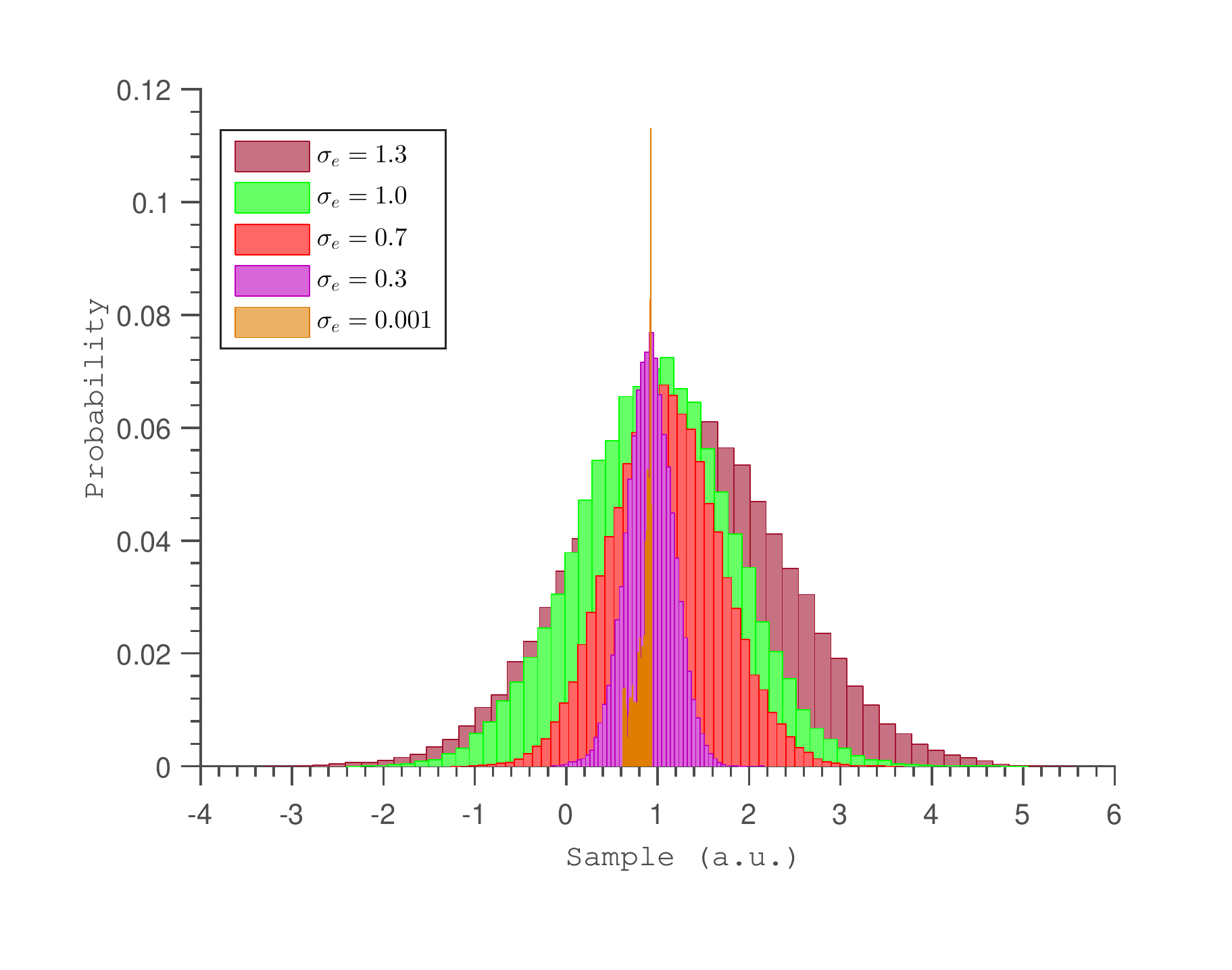}}
\caption{The comparison of marginal probability densities of $f_{v_{1}}$
parameter at all noise levels.}%
\label{fig:v1_hist_overlap_vspace}%
\end{figure}
\begin{figure}[ptb]
\centering
\begin{minipage}{.4\textwidth}
		\centering
		\makebox{\includegraphics[width=1.0\textwidth]{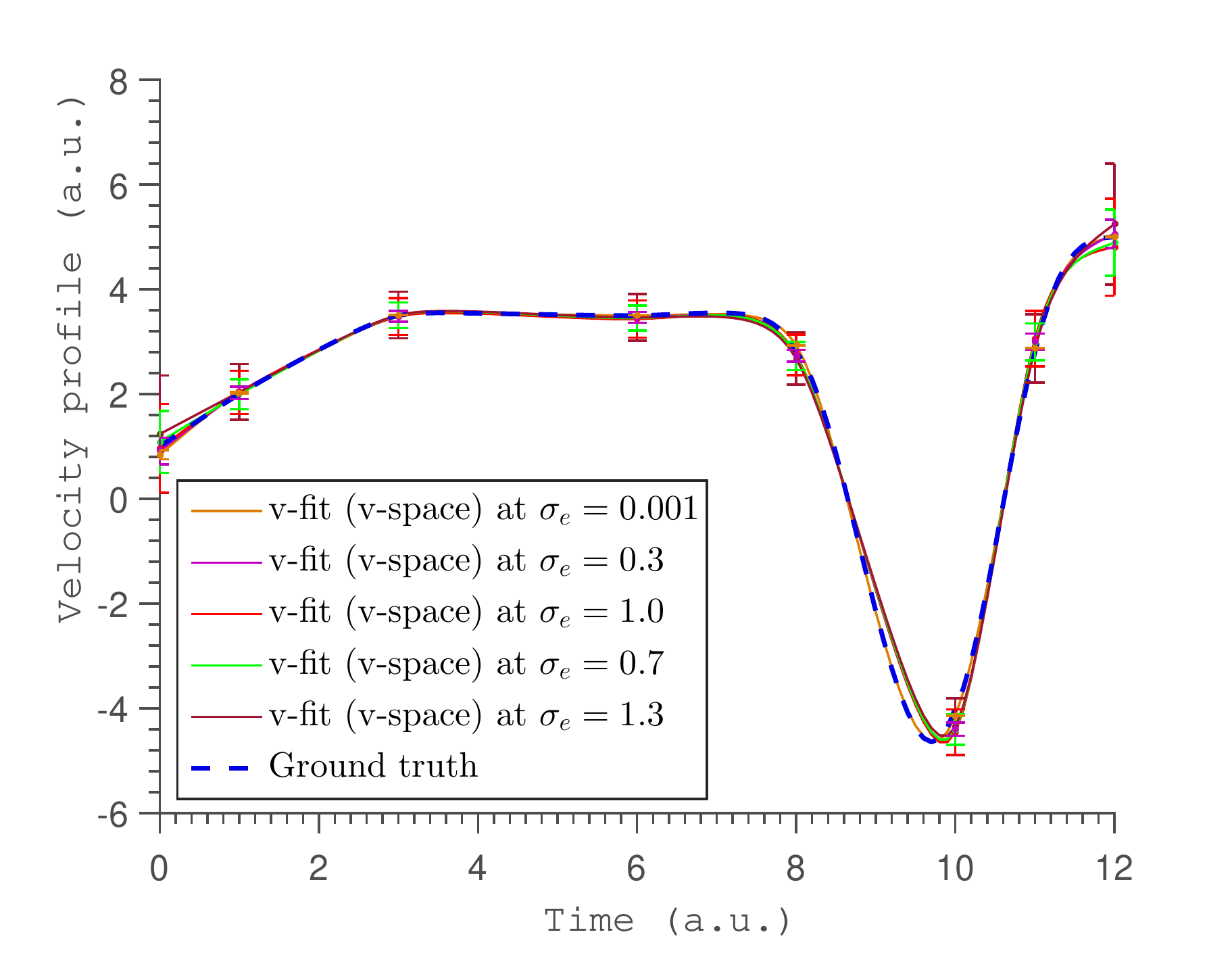}}
		%\label{fig:allNoiselevels_vfit}
	\end{minipage} \begin{minipage}{.4\textwidth}
		\centering
		\makebox{\includegraphics[width=1.0\textwidth]{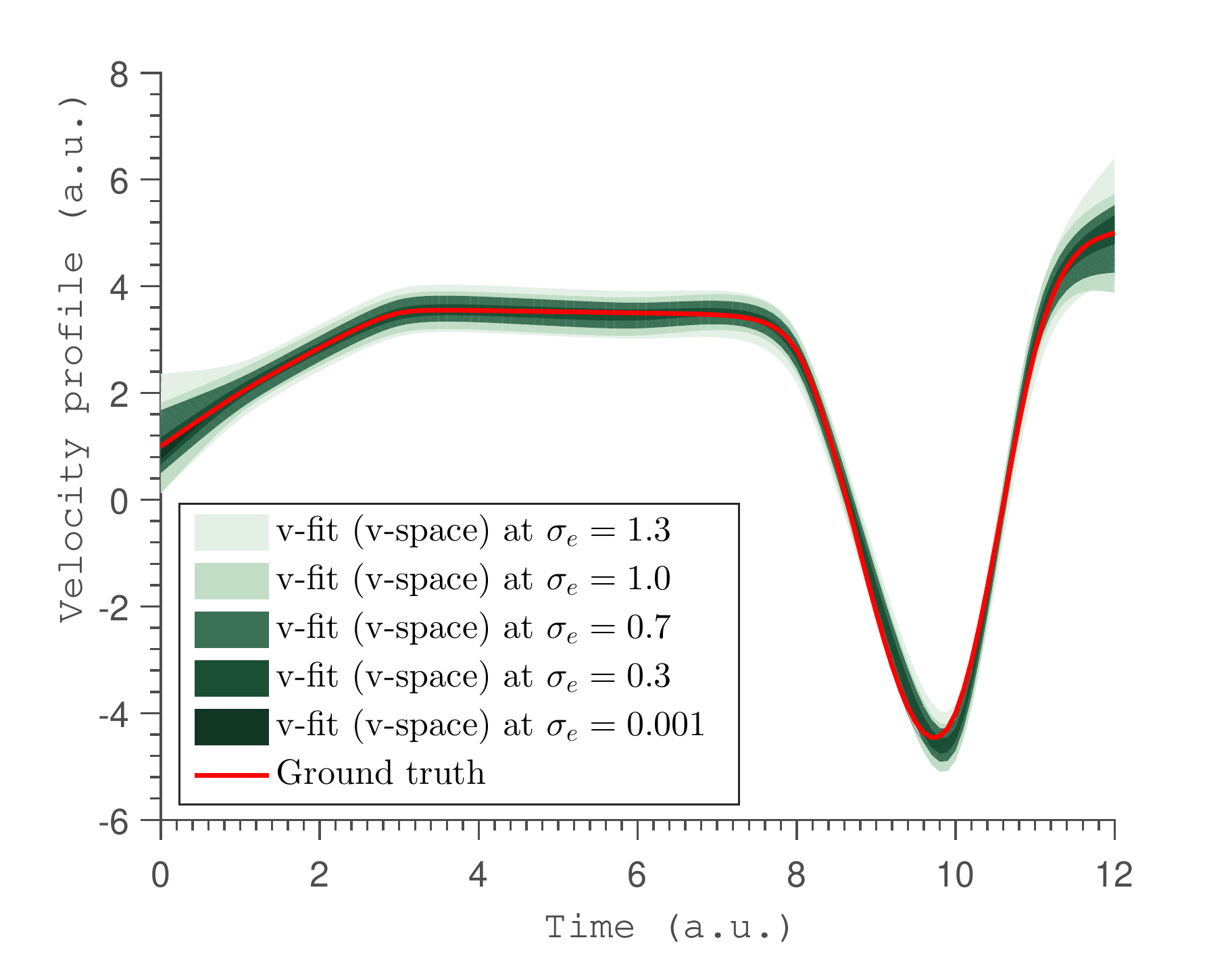}}
		%	\label{fig:sd_all_vfit_vspace_credibles}
	\end{minipage}
\caption{The gradient estimates obtained at the five different noise levels.
The uncertainty of the velocity estimates ($\hat{\sigma}_{v}$) are shown using
error bars (left panel) and credible intervals (right panel).}%
\label{fig:allNoiseLevels_vfit_errorNcredible}%
\end{figure}

 A constant
force is observed between the times $0-8(a.u.)$ and short impulses are
observed between the times $8-11(a.u.)$. Moreover, it is observed that as the
noise level increases, the data around the short impulses tends to get fuzzier
so that it is hard to identify the trajectory. The marginalized
posterior distributions for each parameter is simulated using MCMC sampling.
When the noise increases, the width of the marginal distributions were
expanded. This, in turn, resulted in an increased length of the error
bars of the parameter estimates. The marginal probability distribution of
$f_{v_{1}}$ is shown in Fig. \ref{fig:v1_hist_overlap_vspace}. These
uncertainties of the $f_{v_{i}}$ parameters are shown as error bars
in Fig. \ref{fig:allNoiseLevels_vfit_errorNcredible} (left panel) and as
Bayesian credible intervals in Fig.
\ref{fig:allNoiseLevels_vfit_errorNcredible} (right panel). The gradient
estimate is obtained at all noise levels by placing the model in $\mathbb{V}$
and is shown in Fig. \ref{fig:allNoiseLevels_vfit_errorNcredible}.
The estimates almost overlap with the ground truth data at all noise levels,
except at the boundaries and around the time $t=10a.u.$, where the
short impulses were present in the position profile.

\begin{figure}[ptb]
\centering
\begin{minipage}{.4\textwidth}
		\centering
		\makebox{\includegraphics[width=1.0\linewidth,height=0.7\linewidth]{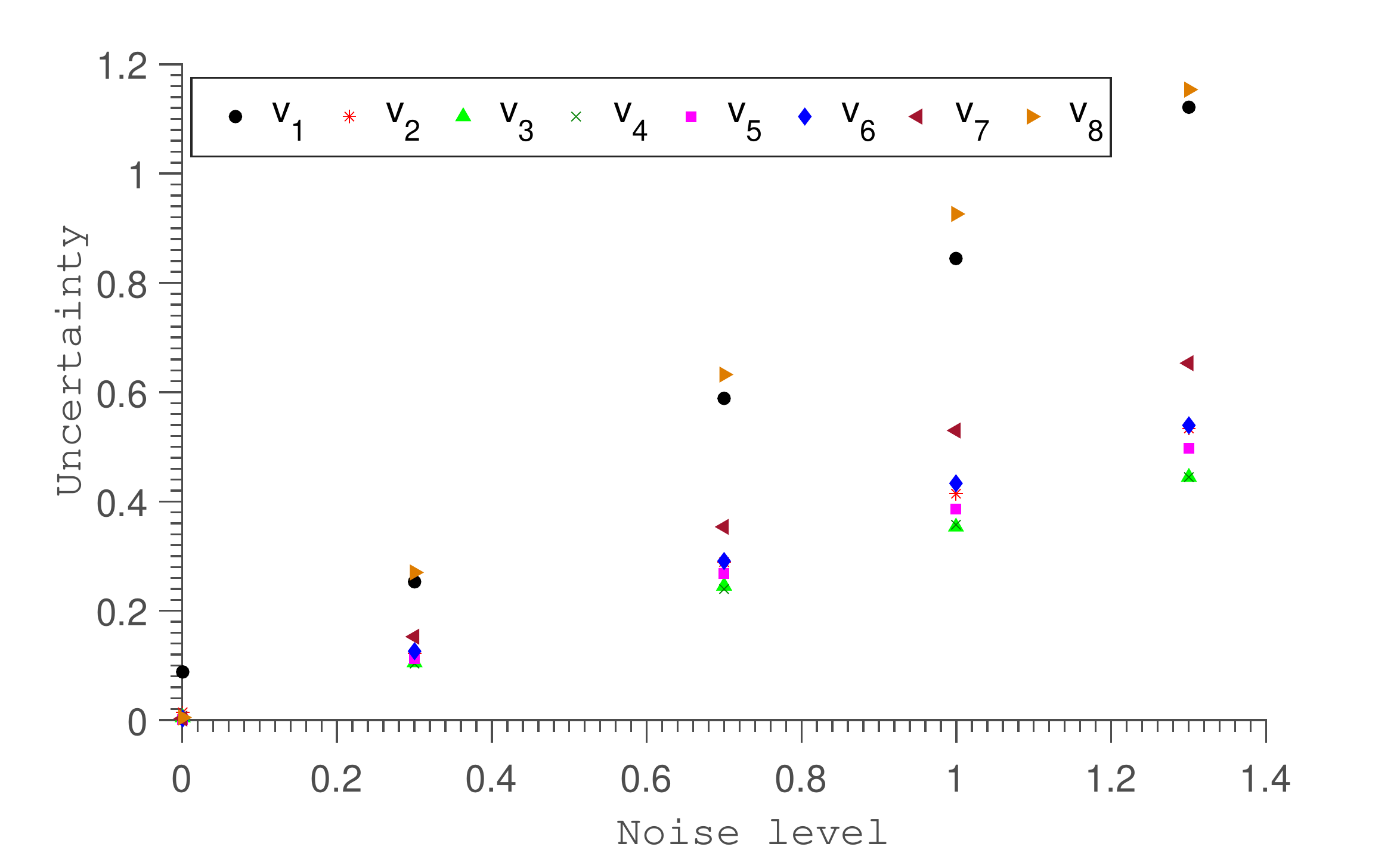}}
		%			\captionof{figure}{A figure}
		\label{fig:errors_v_vfits}
	\end{minipage} \begin{minipage}{.4\textwidth}
		\centering
		\makebox{\includegraphics[width=1.0\linewidth,height=0.7\linewidth]{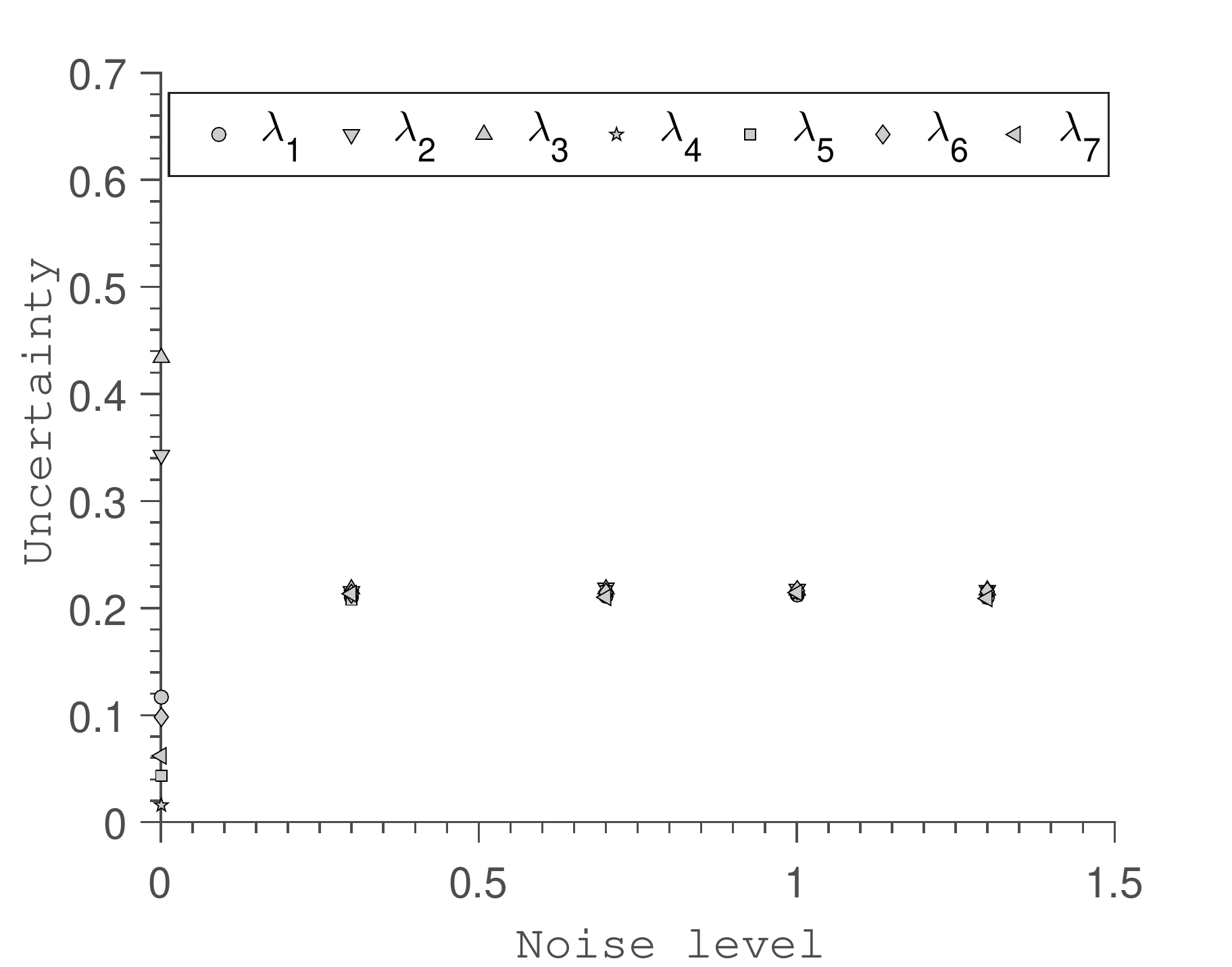}}
		%			\captionof{figure}{Another figure}
		\label{fig:errors_lam_vfits}
	\end{minipage}
\caption{The relationship between uncertainty of the estimates and the noise
level of the system. The left panel depicts the relationship of velocity
parameters whereas the right panel shows the relationship of tension
parameters.}%
\label{all noise levels v fit errors}%
\end{figure}

\begin{figure}[ptb]
\centering
\begin{minipage}{.5\textwidth}
		\centering
		\makebox{\includegraphics[width=1.0\linewidth]{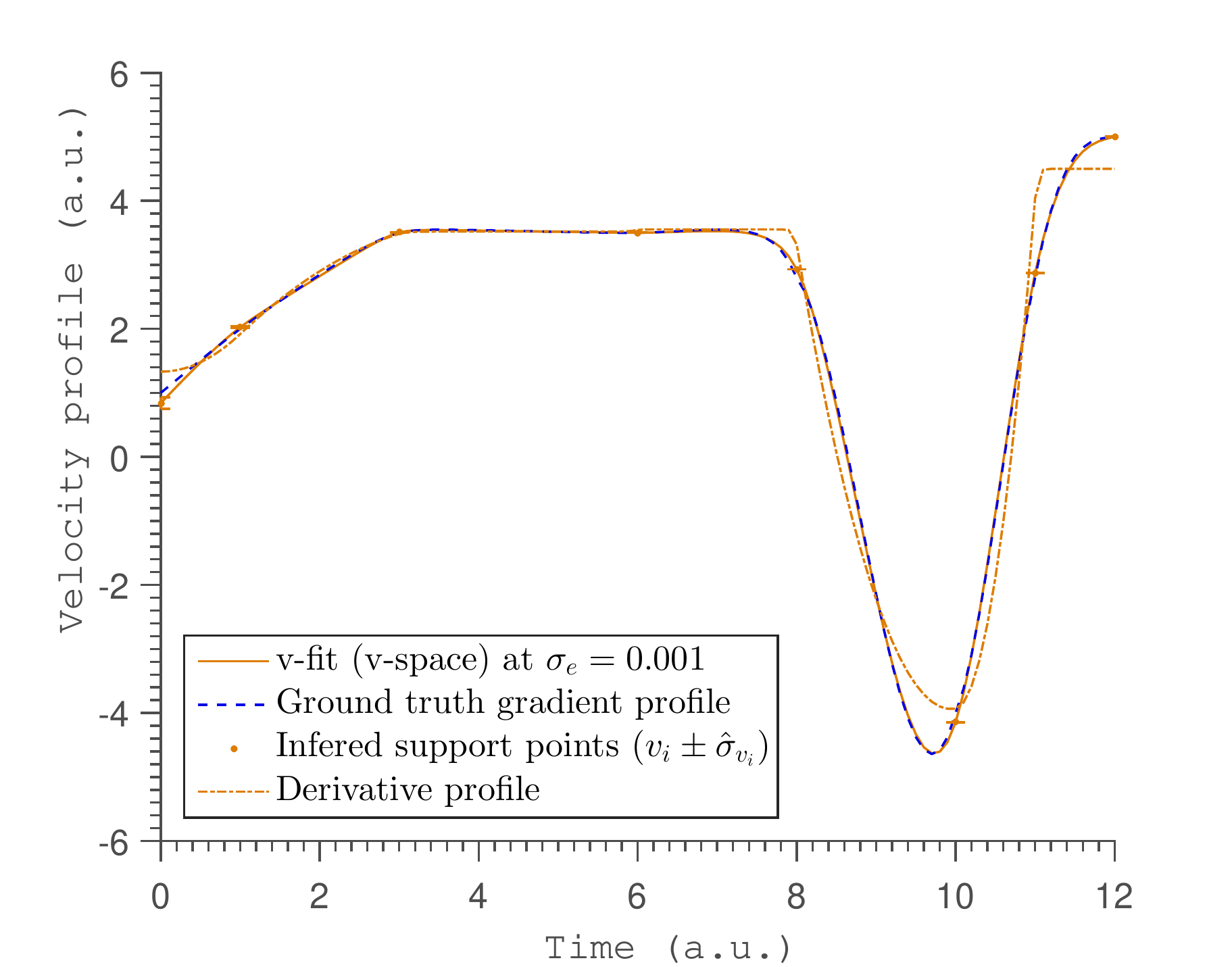}}
		%			\captionof{figure}{A figure}
		\label{fig:sd_0_001_vfit_xv_space}
	\end{minipage}%
 \begin{minipage}{.5\textwidth}
		\centering
		\makebox{\includegraphics[width=1.0\linewidth]{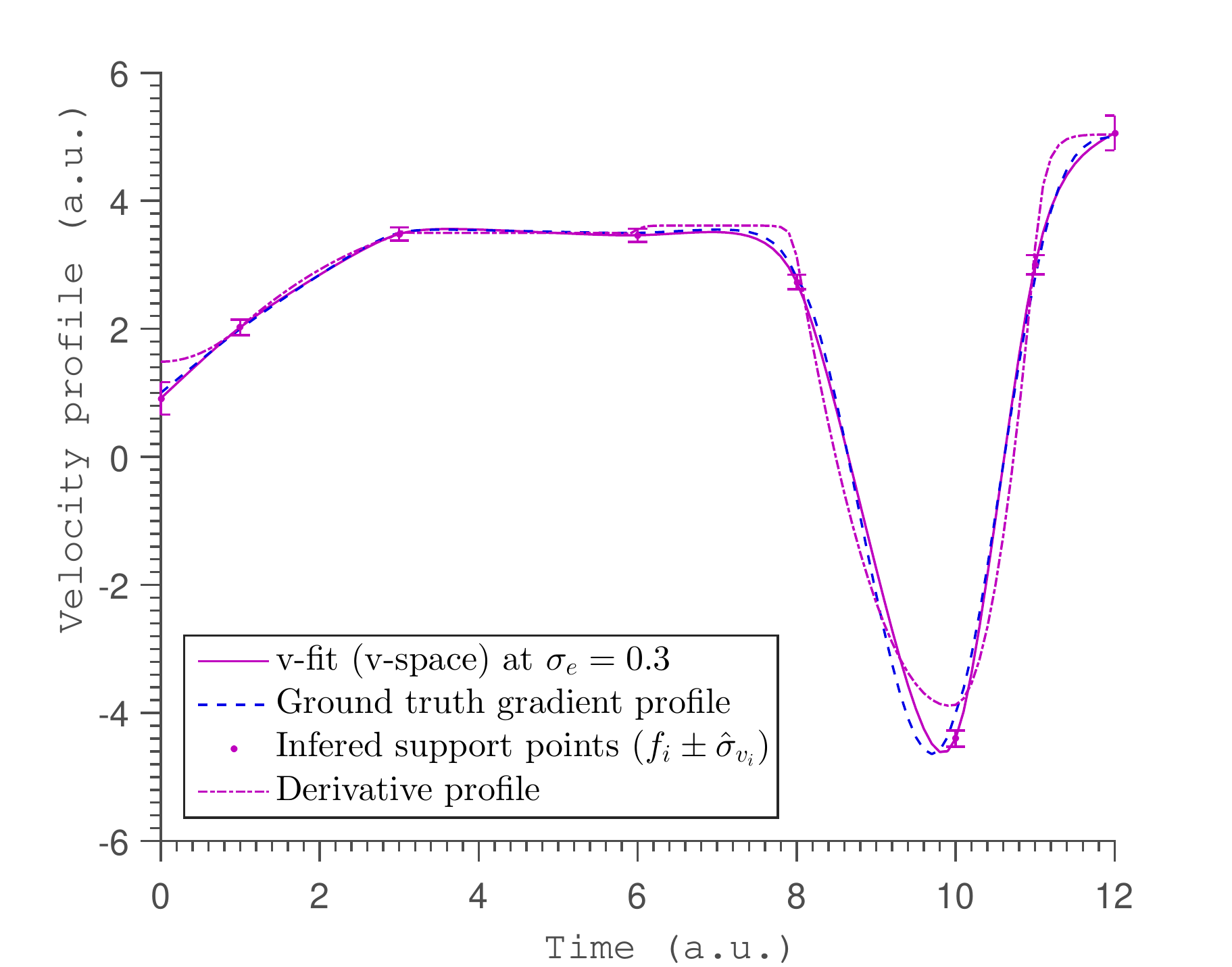}}
		%			\captionof{figure}{Another figure}
		\label{fig:sd_0_3_vfit}
	\end{minipage}
\begin{minipage}{.5\textwidth}
		\centering
		\makebox{\includegraphics[width=1.0\linewidth]{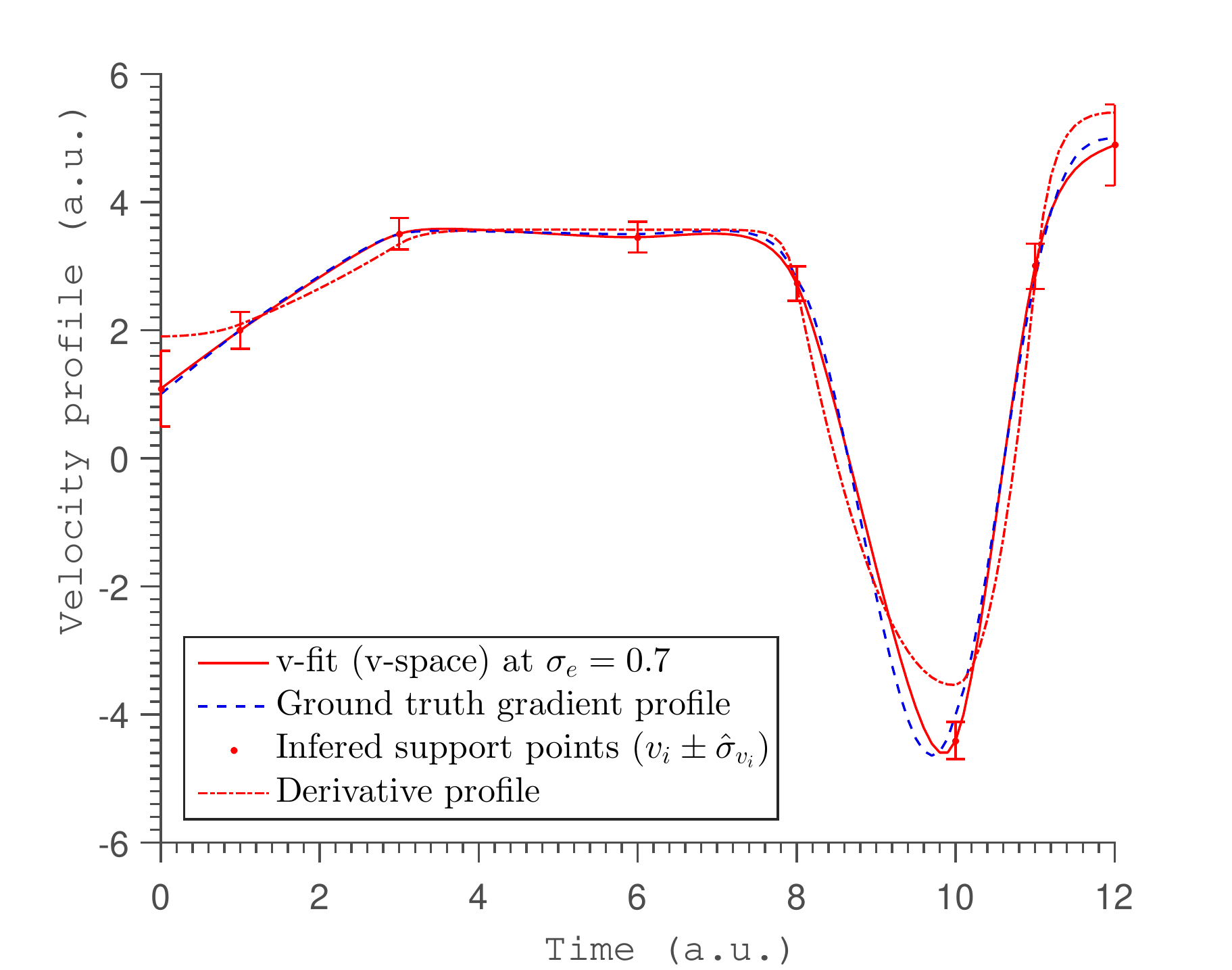}}
		%		\captionof{figure}{A figure}
		\label{fig:sd_0_7_vfit}
	\end{minipage}%
 \begin{minipage}{.5\textwidth}
		\centering
		\makebox{\includegraphics[width=1.0\linewidth]{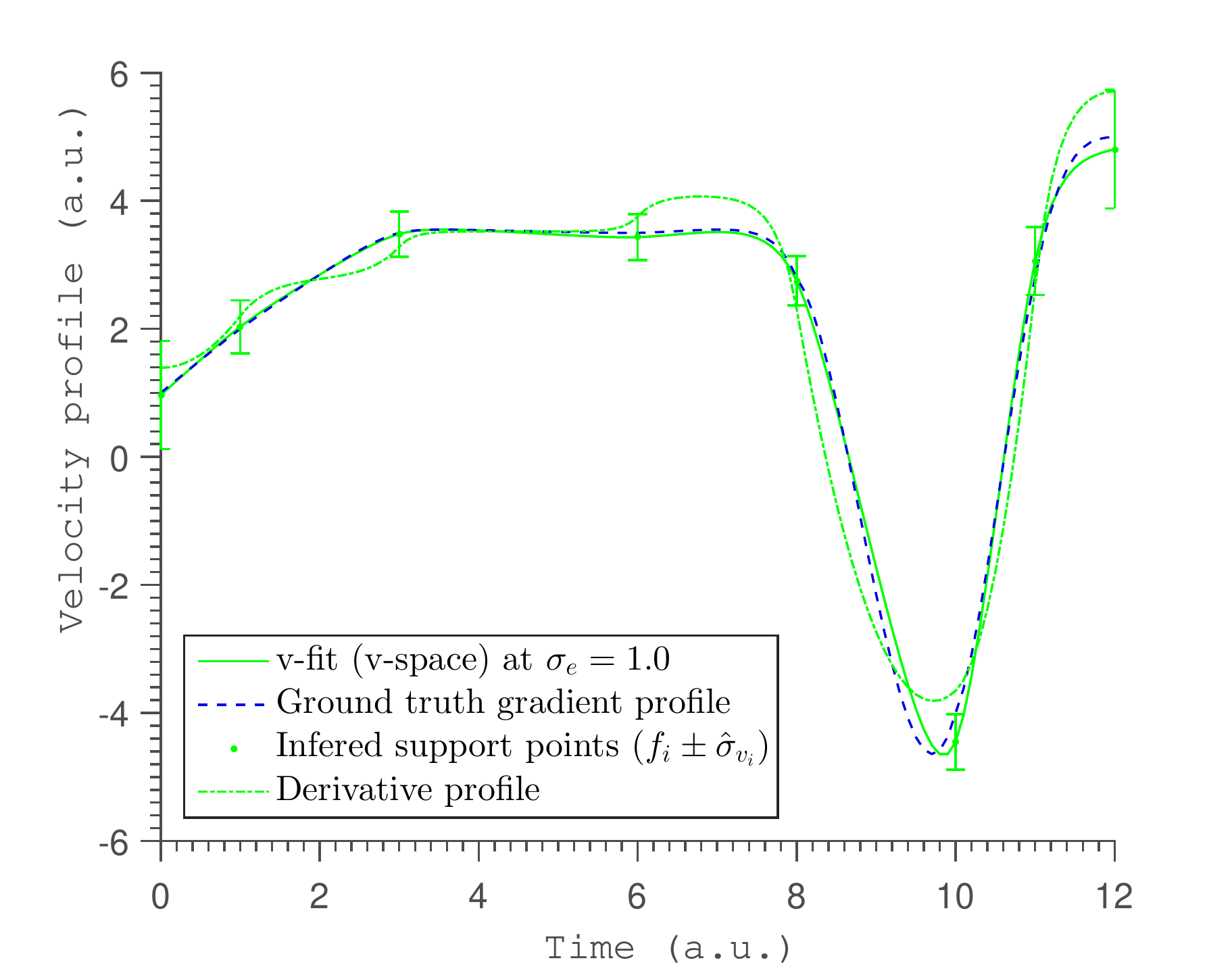}}
		%		\captionof{figure}{Another figure}
		\label{fig:sd_1_0_vfit}
	\end{minipage}
\begin{minipage}{.5\textwidth}
		\centering
		\makebox{\includegraphics[width=1.0\linewidth]{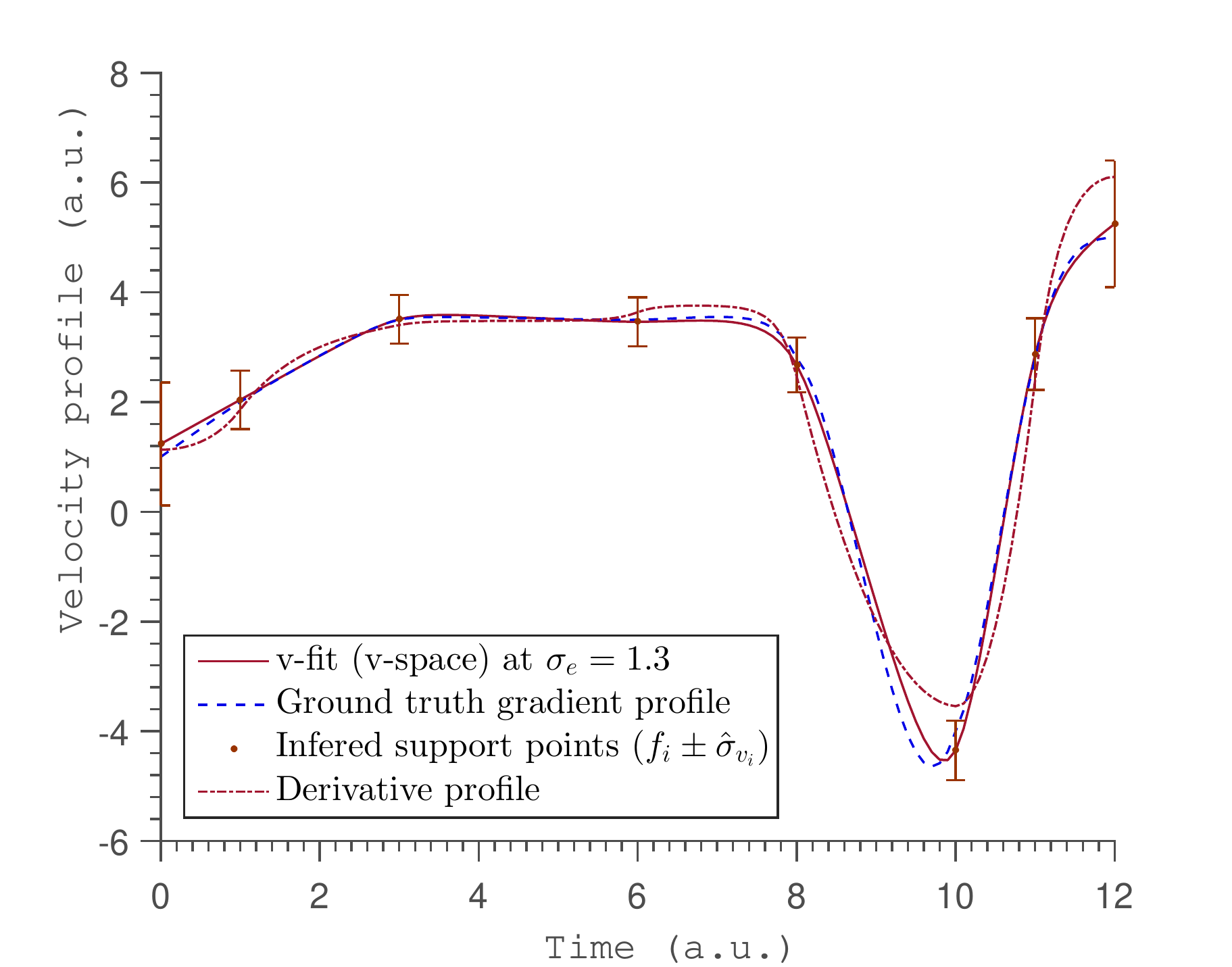}}
		%			\captionof{figure}{Another figure}
		\label{fig:sd_1_3_vfit}
	\end{minipage}
\caption{The panel of velocity estimates from the two spaces compared at five
different noise levels. The noise level increases from left to right and top
to bottom.}%
\label{all noise levels vfit_xv_space}%
\end{figure}

\begin{figure}[ptb]
\centering
\begin{minipage}{.5\textwidth}
		\centering
		\makebox{\includegraphics[width=1.0\linewidth]{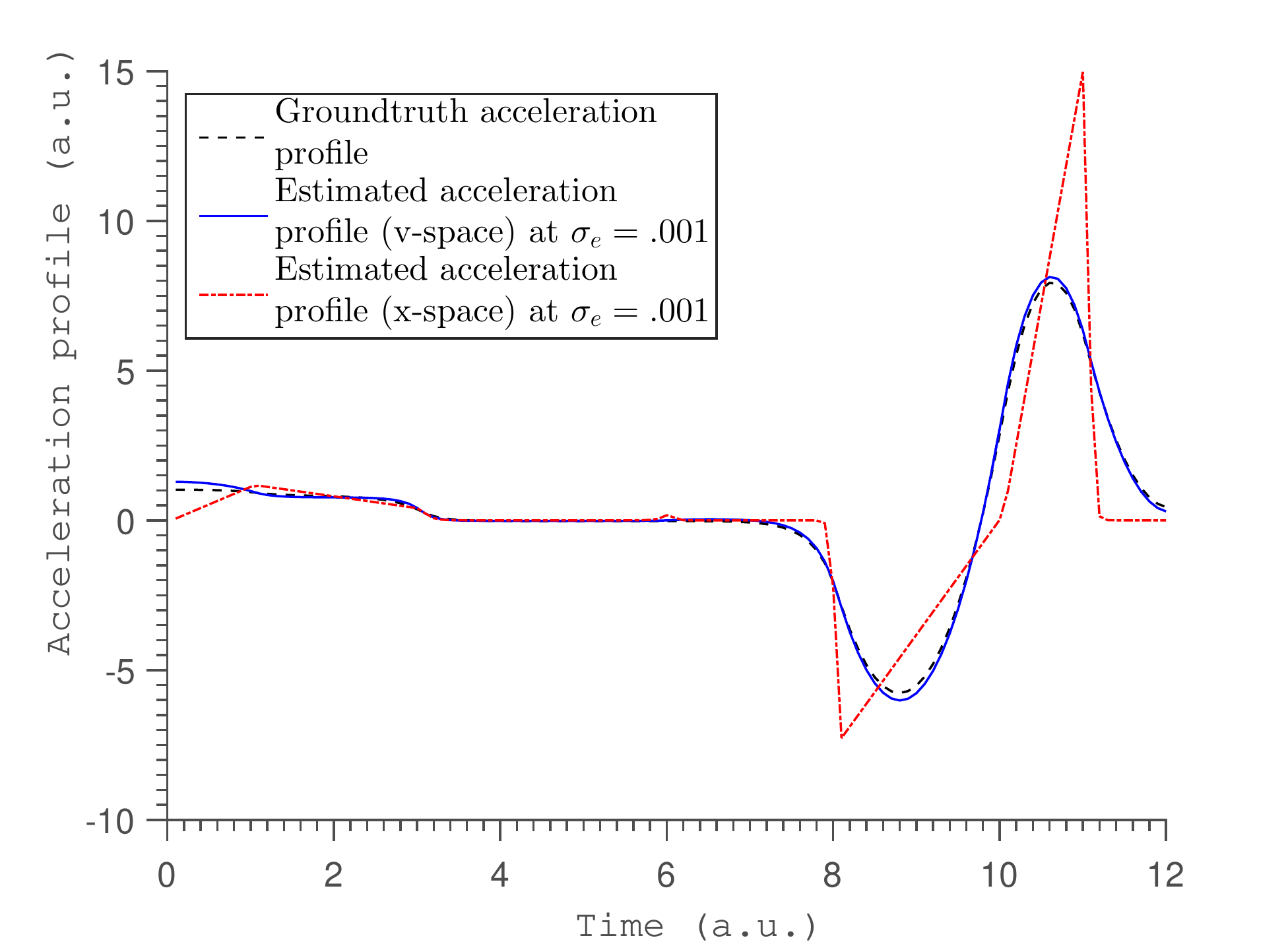}}
		%			\captionof{figure}{A figure}
		\label{fig:sd_0_001_afit_vspace}
	\end{minipage}%
 \begin{minipage}{.5\textwidth}
		\centering
		\makebox{\includegraphics[width=1.0\linewidth]{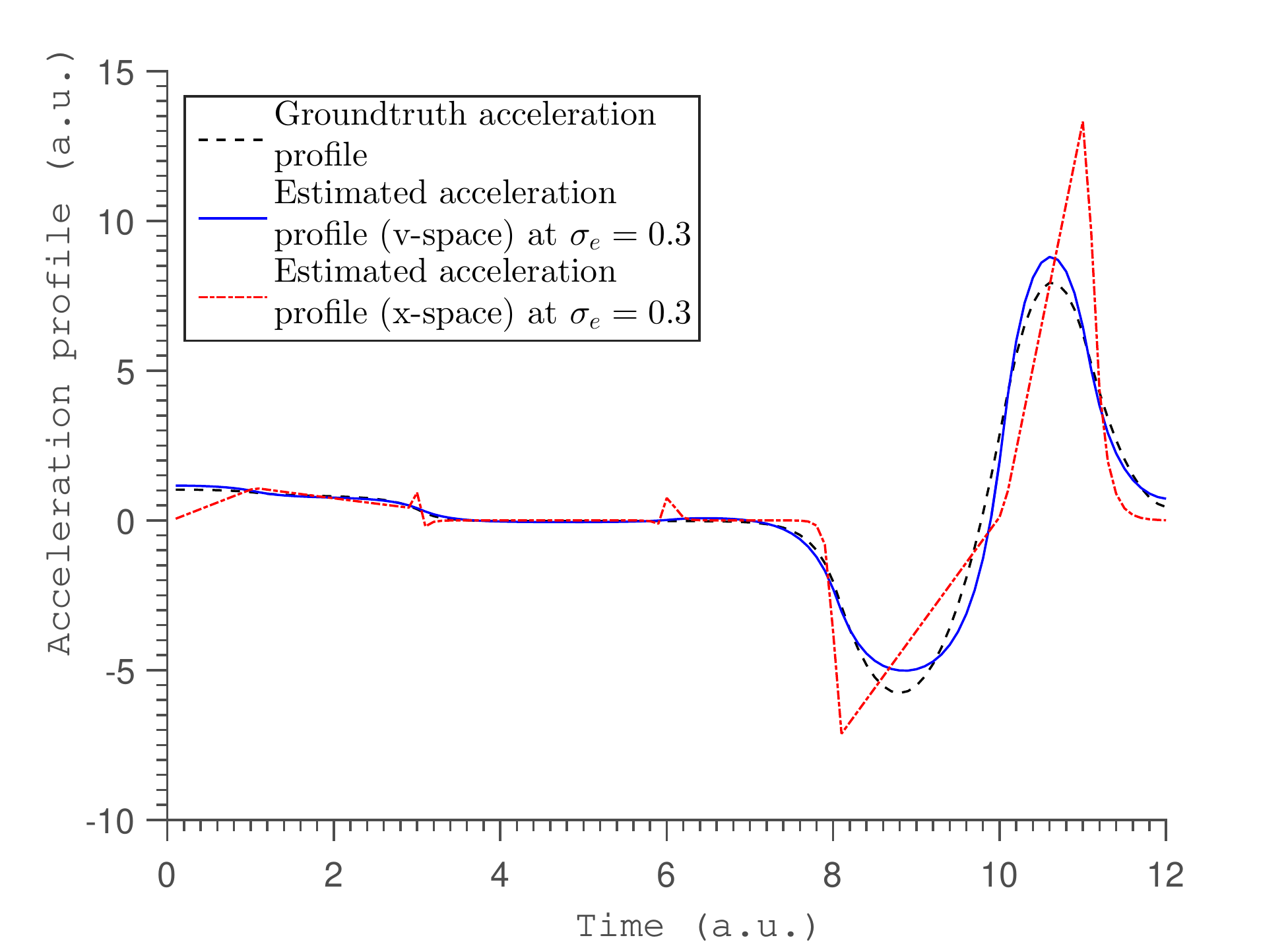}}
		%			\captionof{figure}{Another figure}
		\label{fig:sd_0_3_afit_vspace}
	\end{minipage}
\begin{minipage}{.5\textwidth}
		\centering
		\makebox{\includegraphics[width=1.0\linewidth]{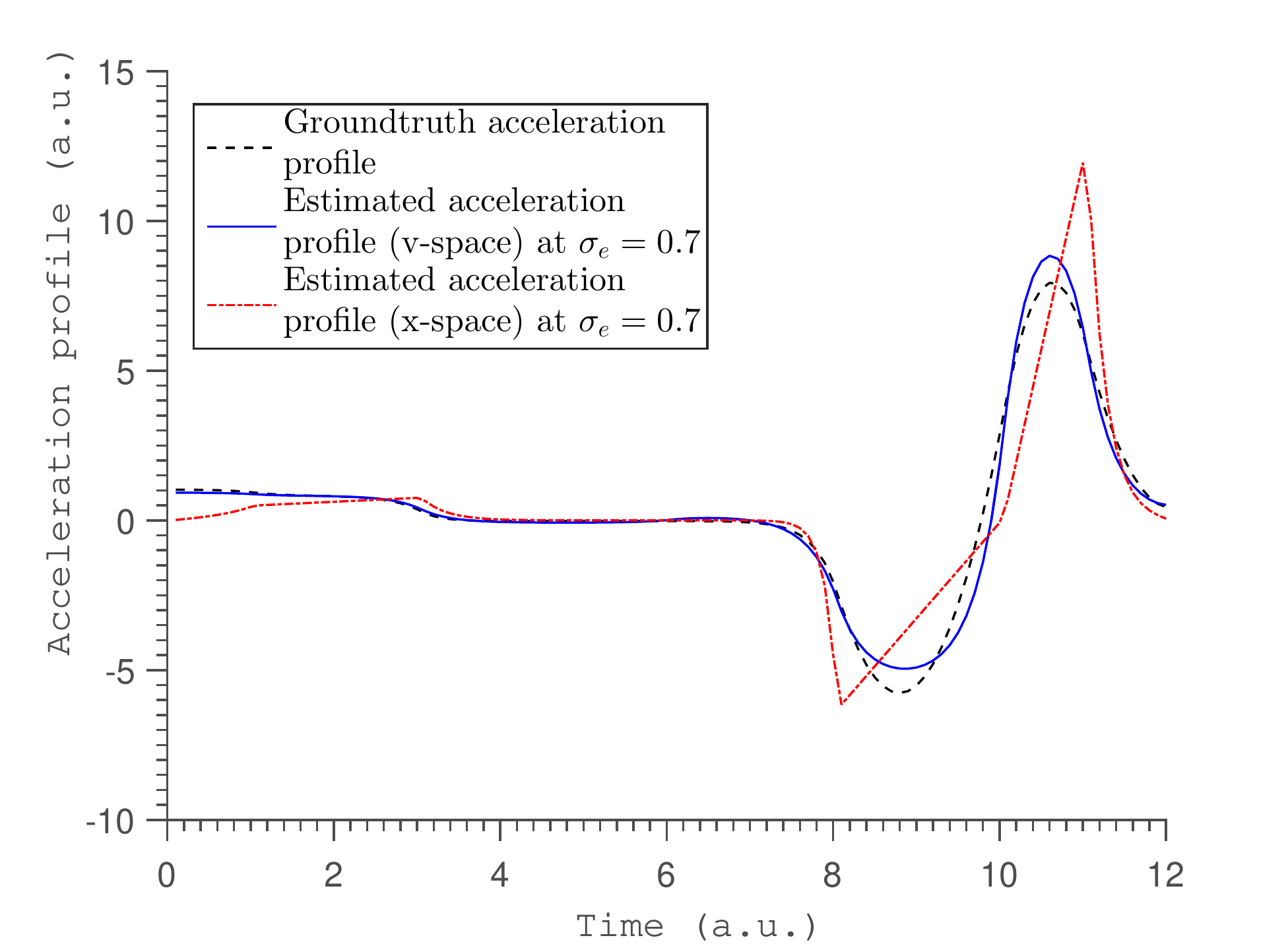}}
		%		\captionof{figure}{A figure}
		\label{fig:sd_0_7_afit_vspace}
	\end{minipage}%
 \begin{minipage}{.5\textwidth}
		\centering
		\makebox{\includegraphics[width=1.0\linewidth]{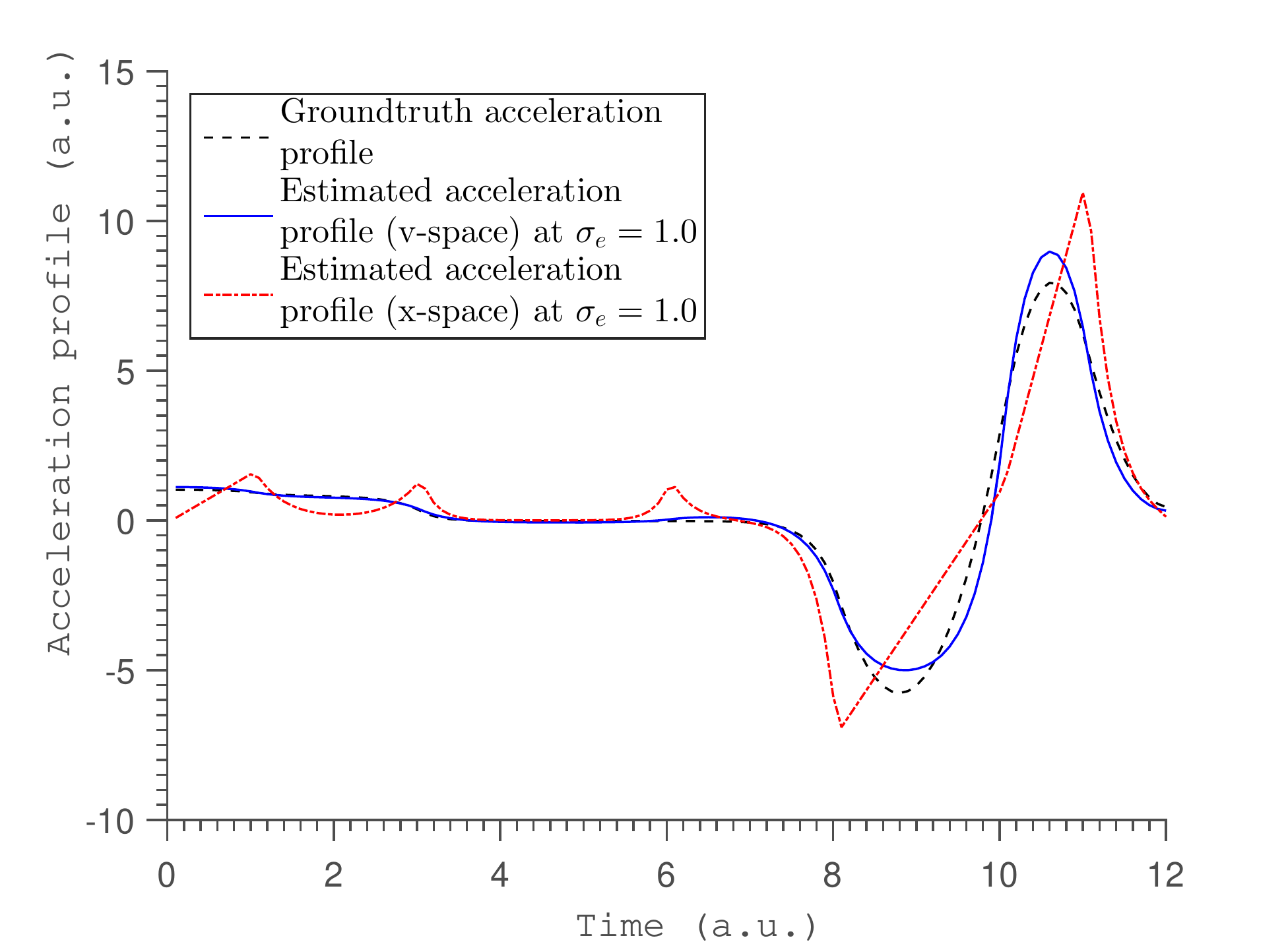}}
		%		\captionof{figure}{Another figure}
		\label{fig:sd_1_0_afit_vspace}
	\end{minipage}
\begin{minipage}{.5\textwidth}
		\centering
		\makebox{\includegraphics[width=1.0\linewidth]{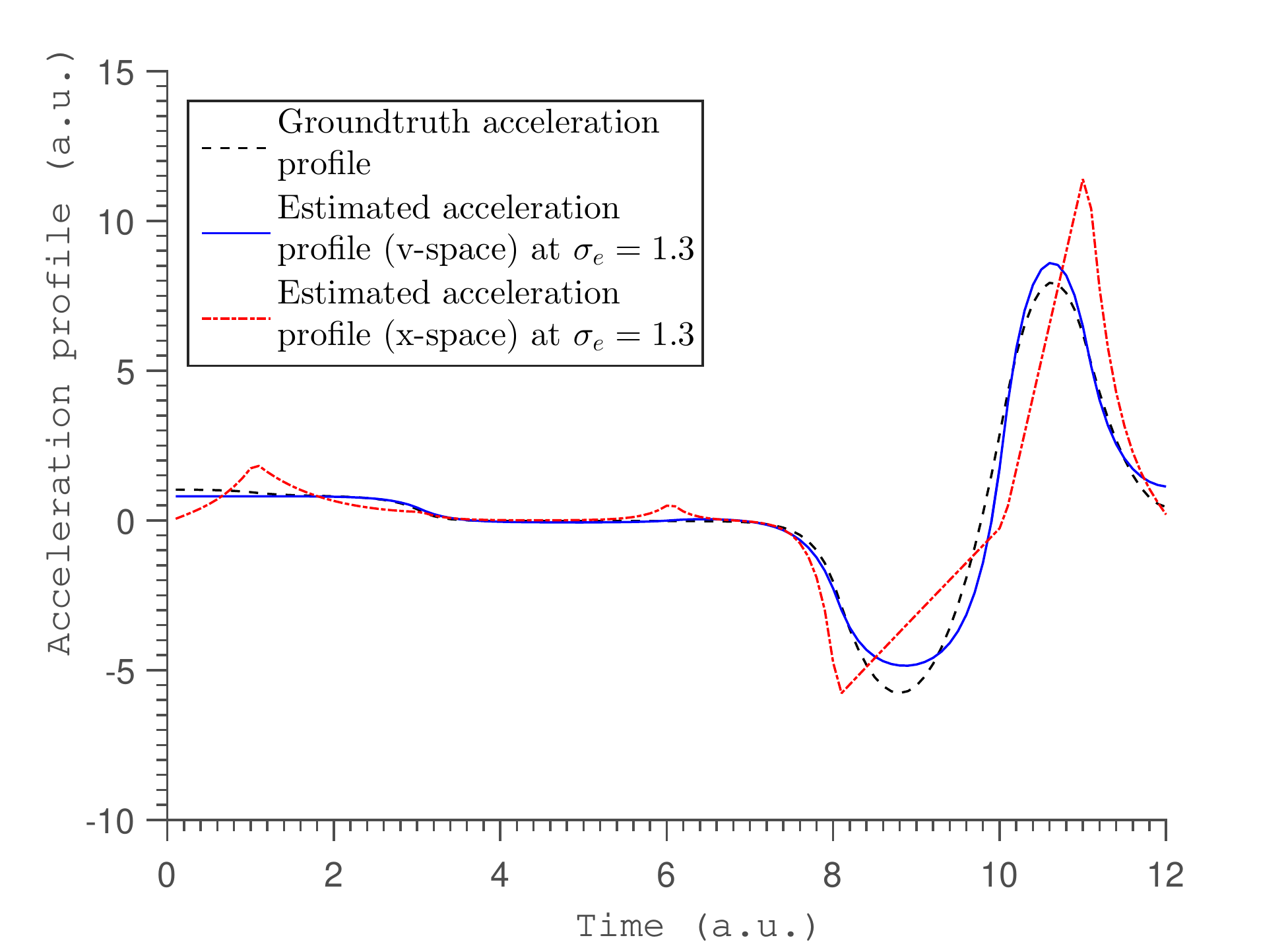}}
		%			\captionof{figure}{Another figure}
		\label{fig:sd_1_3_afit_vspace}
	\end{minipage}
\caption{The panel of acceleration estimates obtained from two methods are
compared at five different noise levels.}%
\label{all noise levels afit_xv_space}%
\end{figure}

\begin{figure}[ptb]
\centering
\begin{minipage}{.5\textwidth}
		\centering
		\makebox{\includegraphics[width=1.0\linewidth]{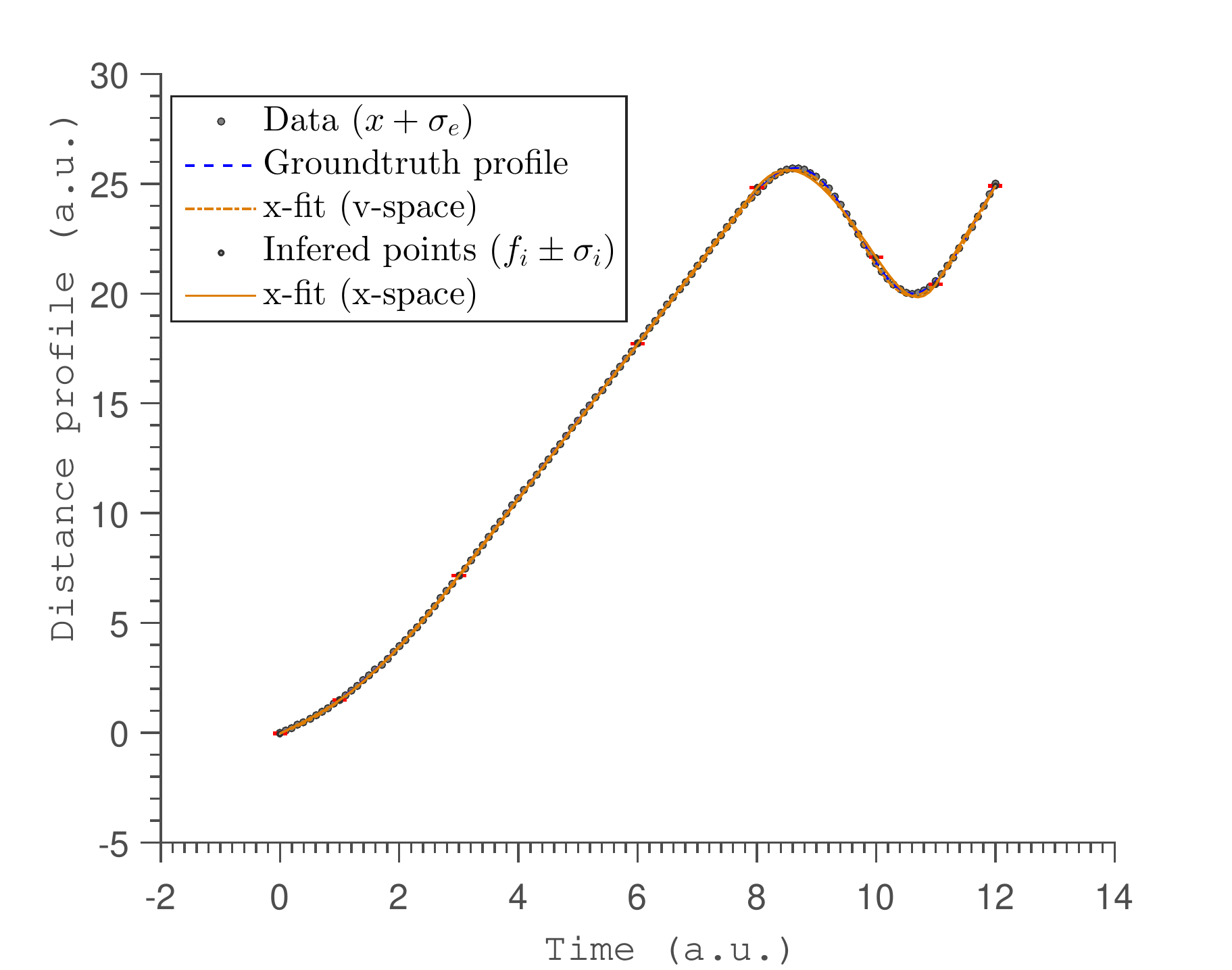}}
		%		\captionof{figure}{A figure}
		\label{fig:sd_0_001_xfit_xv_space}
	\end{minipage}%
 \begin{minipage}{.5\textwidth}
		\centering
		\makebox{\includegraphics[width=1.0\linewidth]{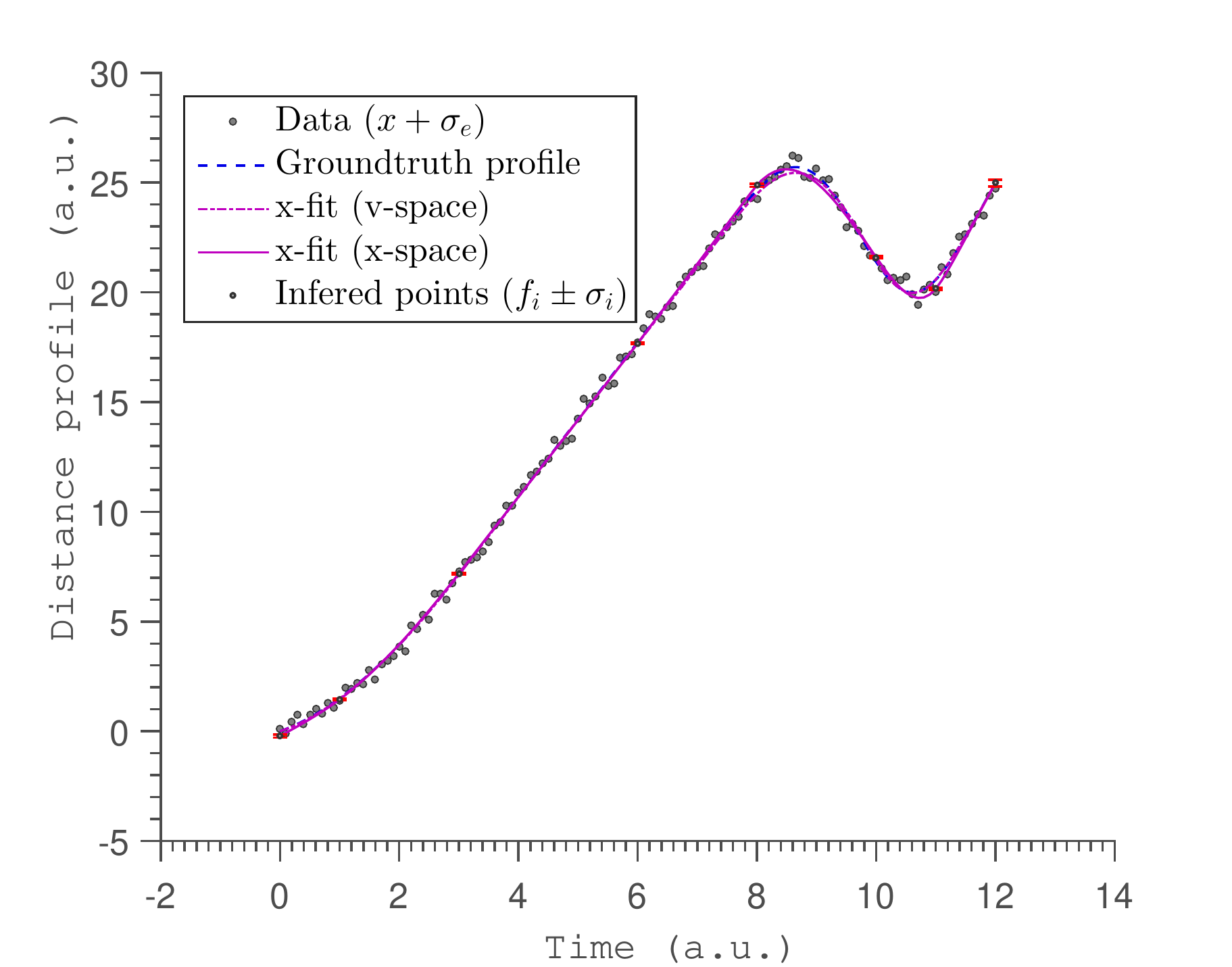}}
		%		\captionof{figure}{Another figure}
		\label{fig:sd_0_3_xfit_xv_space}
	\end{minipage}
\begin{minipage}{.5\textwidth}
		\centering
		\makebox{\includegraphics[width=1.0\linewidth]{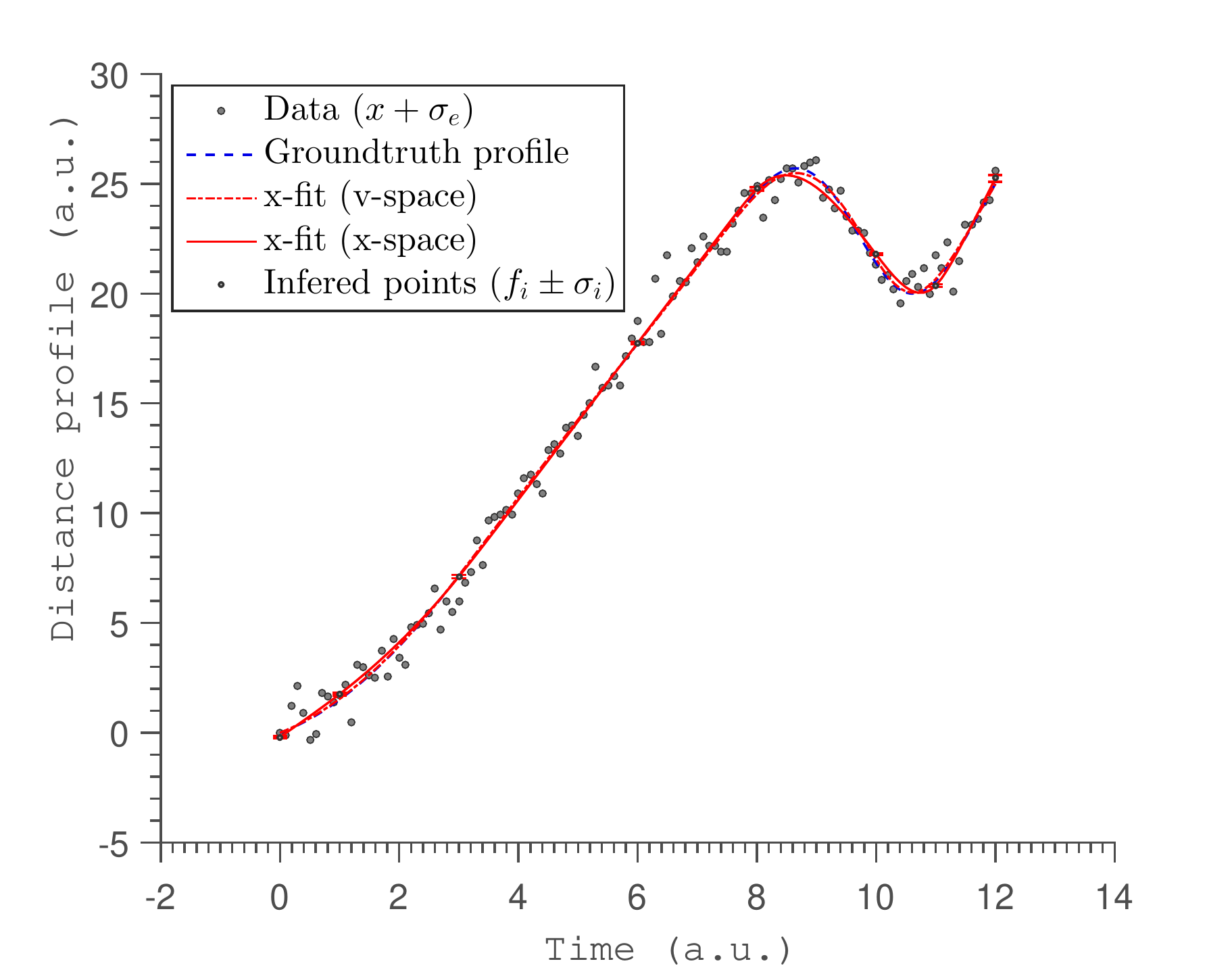}}
		%		\captionof{figure}{A figure}
		\label{fig:sd_0_7_xfit_xv_space}
	\end{minipage}% 
\begin{minipage}{.5\textwidth}
		\centering
		\makebox{\includegraphics[width=1.0\linewidth]{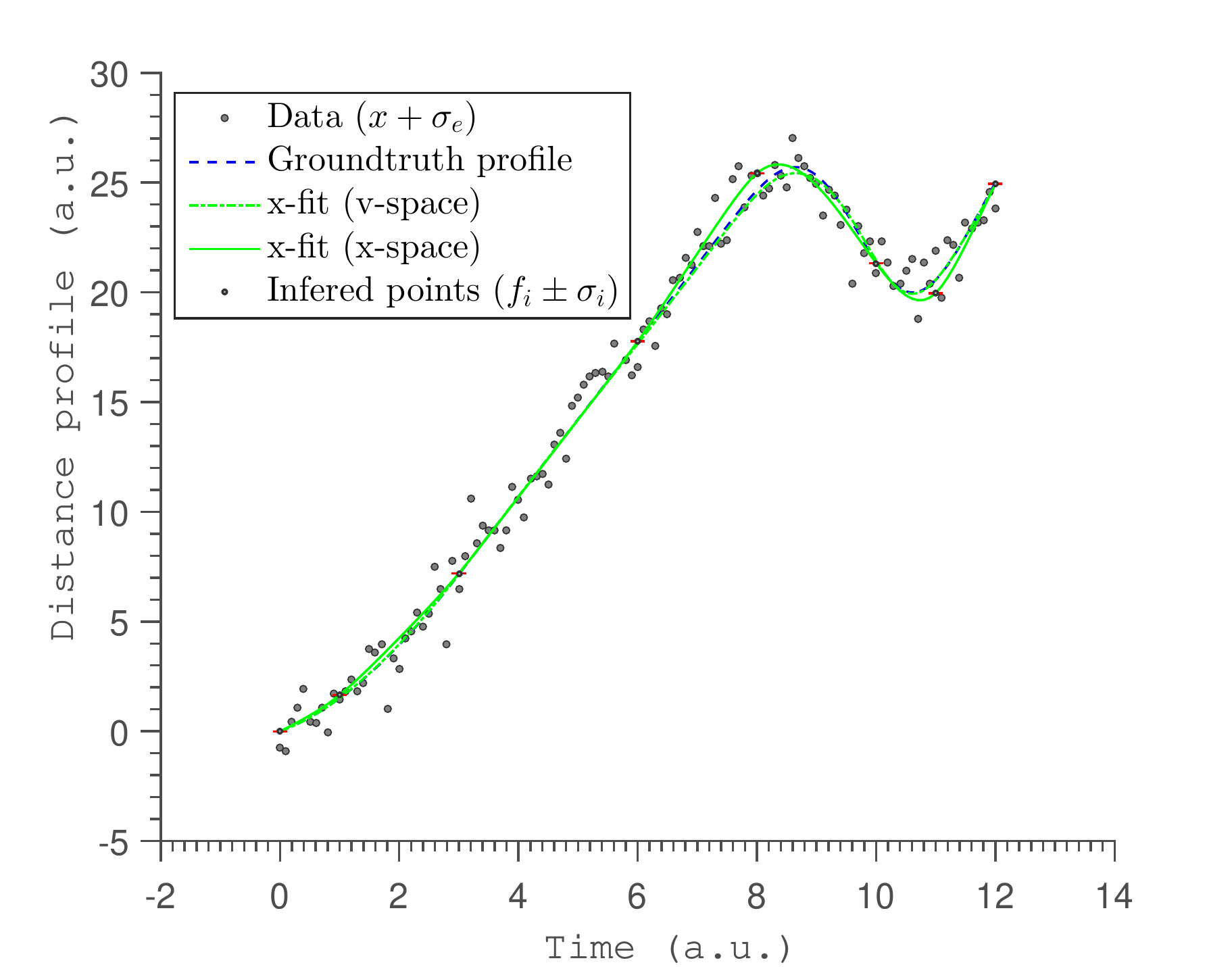}}
		%		\captionof{figure}{Another figure}
		\label{fig:sd_1_0_xfit_xv_space}
	\end{minipage}
\begin{minipage}{.5\textwidth}
		\centering
		\makebox{\includegraphics[width=1.0\linewidth]{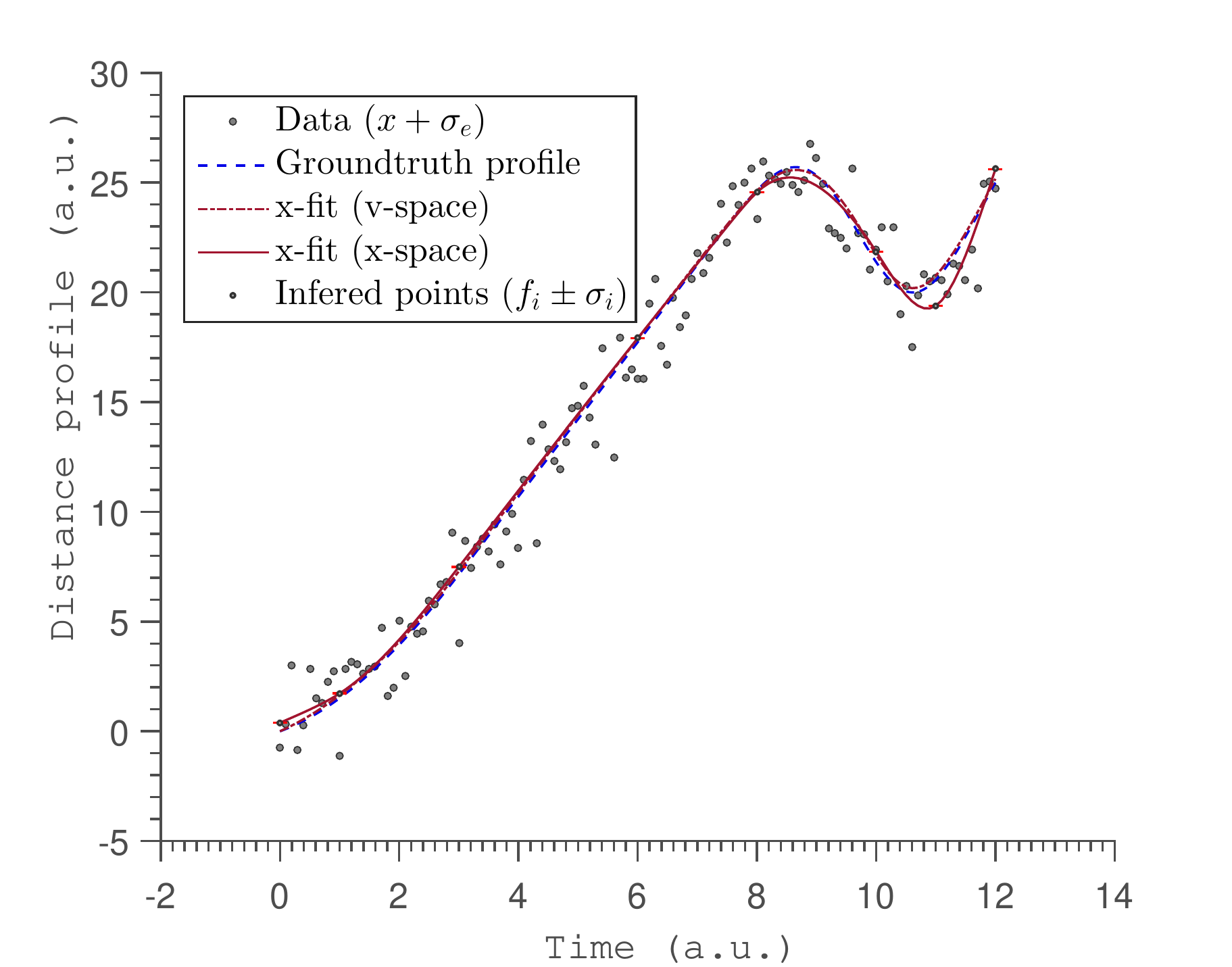}}
		%			\captionof{figure}{Another figure}
		\label{fig:sd_1_3_xfit_xv_space}
	\end{minipage}
\caption{The panel of x-fit estimates obtained from two different spaces
compared at five different noise levels.}%
\label{all noise levelsxfit_xv_space}%
\end{figure}

The uncertainties of both velocity and tension parameters are
characterized in Fig. \ref{all noise levels v fit errors}. The uncertainty of
the velocity parameters with the noise level exhibits a linear
relationship reflecting the linear relationship of velocity parameters in the
likelihood function. The uncertainty of the tension parameters does not show
any specific pattern as it has a non linear pattern in the
likelihood.
The gradient estimates are compared with the analytic derivative
of the x-fit obtained in $\mathbb{U}$. The x-fit is obtained by fitting the
same exponential cubic spline to the time-position data in $\mathbb{U}$. This
was performed at the same knot positions. The gradient profile, the
acceleration profile, and position profile were obtained from the two methods
(via v-fit from $\mathbb{V}$ and x-fit from $\mathbb{U}$) and compared in Fig.
\ref{all noise levels vfit_xv_space}, Fig.
\ref{all noise levels afit_xv_space} and Fig.
\ref{all noise levelsxfit_xv_space}, respectively. The two
types of estimates of acceleration profile from the v-space and the x-space
were obtained by differentiating the v-fit (v-space) and by finite
differencing the v-fit (x-space), respectively, from Fig.
\ref{all noise levels vfit_xv_space}.
%For further reference of estimators, uncertainties, MCMC sample chains and auto-correlations are given in Appendix C.

The gradient estimates that emerge from the two spaces
(namely, the x-space and v-space) clearly show some remarkable
differences. In particular, the difference in accuracy is shown in Table
\ref{error norm v fit x fit} using the error norm,
\begin{equation}
||e||^{2}\overset{\text{def}}{=}\sum_{i=1}^{L}\left(  f_{v_{\text{estimated}%
_{i}}}-f_{v_{\text{true}_{i}}}\right)  ^{2}\text{,}%
\end{equation}
where $L$ is the sample size \cite{hansen1992analysis}.
\begin{table}[ptb]
\caption{The comparison of accuracy of velocity fit from v space, and x
space.}%
\label{error norm v fit x fit}%
\centering
\fbox{
\begin{tabular}{l|l|l}
& \multicolumn{2}{c}{$||e||^{2}$}\\ \hline
noise level & v-fit    & \begin{tabular}[c]{@{}c@{}}Derivative \\ of x-fit\end{tabular}\\ \hline
$0.001$     & $0.2077$ & $14.4107$\\
$0.3$       & $1.9679$ & $15.0792$\\
$0.7$       & $3.1793$ & $20.9974$\\
$1$         & $3.1569$ & $30.1122$\\
$1.3$       & $3.8159$ & $31.2719$%
\end{tabular}
}
\end{table}

The acceleration characterizes the acting forces on the
object. The acceleration estimate (x-space) could not achieve constant force
where it is expected. The errors at short impulses around $t=10$ a.u. started
getting very large and this indicates the possibility of infinite forces.
Moreover, the acceleration profile from the x-space (red dashed-dot
lines) is not realistic with its very sharp turns. Therefore, the
acceleration estimate from the velocity space is a reliable estimator of
acting forces.  The goodness of the fit is tested by studying the
error sum of squares (\textrm{SSE}) defined as \cite{merriman1909text},%
\begin{equation}
\mathrm{SSE}\overset{\text{def}}{=}\sum_{i=1}^{n}\left(  \hat{f}_{i}%
-f_{i}\right)  ^{2} \label{eq: sse}%
\end{equation}
where $\hat{f}_{i}$ represent the estimated function values.
The quantity $f_{i}$ in Eq. ~\eqref{eq: sse} is replaced with noisy
$f_{i}$ to obtain the second and the third columns in Table
\ref{all noise sse comparison} while $f_{i}$ is replaced with
the original $f_{i}$ without noise to calculate the fourth
and fifth columns in the same table. The trajectory estimate is a bonus
product from the gradient profile (v-space) and is obtained by integrating the
gradient estimate (v-space). The trajectory estimates (position profiles) are
compared in Fig. \ref{all noise levelsxfit_xv_space}. The x-fit (x-space)
follows the noisy data closely. This is because when fitting a curve in the
same space as data, that curve is trying to minimize the SSE which means it
tries to match the data as much as possible. Therefore, the x-fit from
$\mathbb{U}$ tries to satisfy the noisy data as much as possible. \ However,
the x-fit (v-space) follows the ground truth data more than it follows the
noisy data. When integrating the v-fit (v-space), an additional order of
differentiability is added to the trajectory estimate. Therefore, the
 x-fit (v-space) has a higher order of differentiability than that
of the x-fit (x-space). This additional order of differentiability ensures the
continuity of the velocity and, therefore, it satisfies the
constraint of the finite force of the object.
\begin{table}[ptb]
\caption{The comparison of the quality of x-fit from x and v spaces
respectively using error sum of squares (SSE).}%
\label{all noise sse comparison}%
\centering
\fbox{
\begin{tabular}
[c]{l|ll|ll}
& \multicolumn{2}{l|}{SSE with noisy data} & \multicolumn{2}{l}{SSE with true
data}\\\hline
$\sigma_{e}$ & x-fit ($\mathbb{U}$) & x-fit ($\mathbb{V}$) & x-fit
($\mathbb{U}$) & x-fit ($\mathbb{V}$)\\\hline
$0.001$ & $1.1743$ & $0.6082$ & $1.1753$ & $0.6089$\\
$0.3$ & $12.4677$ & $12.6049$ & $2.1946$ & $0.8541$\\
$0.7$ & $62.4582$ & $63.4059$ & $4.1855$ & $0.8176$\\
$1.0$ & $121.0864$ & $120.8161$ & $13.3478$ & $1.1253$\\
$1.3$ & $163.1339$ & $264.3176$ & $18.9582$ & $3.1651$%
\end{tabular}
}\end{table}The calculated squared sum of errors (SSE) with true data and
noisy data are shown in the Table \ref{all noise sse comparison}. The SSE
values for noisy data are smaller for the x-fit (x-space).
This, in turn, reflects the fact that the x-fit (x-space) follows
noisy data better. However, the SSE values for true data are smaller
for the x-fit (v-space). This fact, instead, reflects the
fact that the x-fit (v-space) follows true data better.

\section{Conclusion}

In this paper, we proposed a method to compute the profile gradient
of a noisy system using Bayesian inference strategy.
The Bayesian inference method allowed the inference of an
un-observable quantity, the gradient (velocity), by building a meaningful
relationship between the un-observable gradient profile in one space
(that is, the velocity space) and the observable data
profile in another space (that is, the data space - $\mathbb{U}$). Furthermore, the
gradient profile was modeled by the exponential cubic spline. The results of this
new method are compared against a common method of fitting an
exponential cubic spline in data space ($\mathbb{U}$) and taking the
subsequent analytic derivative. The fitting in $\mathbb{U}$ was also performed in a
Bayesian framework and the same DRAM sampler was used to simulate the
posterior distributions. The results show that the gradient estimates obtained
by modeling in v-space produce more accurate estimates than the
results of the traditional method (see Table III). Moreover, it was able to produce
better acceleration estimates with reliable and accurate values where a
constant force is expected (see FIG. 7). The third type of estimate, the trajectory
profile was also compared. We conclude
that the x-fit estimates follow noisy data when solved by placing the model in the
x-space while the x-fit estimates follows the ground truth data when solved placing the model in the velocity space (see FIG. 8 and Table IV). It is argued that by integrating the model
in velocity space, an extra order of differentiability is added to x-fit ensuring the finite force constraint of a moving object.

In conclusion, the method demonstrated in this paper is a superior
method to estimate the velocity of a moving object under finite force as
compared to others in literature (for instance, see
\cite{von2011digital} and references therein). It ensured better estimates in
all three cases: trajectory, velocity, and the acceleration.

We point out that although our main focus in this paper is on the proposal
of a novel theoretical method to estimate a profile gradient from discrete
noisy data, a number of improvements can be sought in its practical
implementation. For instance, the use of an MCMC algorithm gains precision
at the expense of speed. A faster algorithm (such as the Expectation
Maximization algorithm\ \cite{dempster77})\ may be needed for real time
estimation. Furthermore, the amount and placement of the knots lacks a
systematic guiding principle. In the future we plan to use Bayesian model
selection for determining the amount. For the placement issue, we can adopt
a hierarchical approach by including spline knot location algorithms \cite%
{montoya14} in conjunction with our main algorithm. This would
provide estimates for the values associated with the knots and
where they should be located as well. Finally, we hope to apply our Bayesian
estimation technique to more realistic problems where acceleration,
velocity, and trajectory estimations are needed.

%\begin{acknowledgments}
The authors would like to thank the meaningful and helpful discussions had
with Udo Von Toussaint at the Max-Planck Institute in Garching,
Germany.
%\end{acknowledgments}

\pagebreak

\appendix

\section{} \label{appendix I}

In this Appendix, we present some details on the integration
of the exponential cubic spline. The integration is done in two cases.
In one case, one integrates the spline function between two knots and it is
referred to as $I_{\text{full}}$. In the other case, one integrates
the spline function from a knot to a random point before the next knot and it
is referred to as $I_{\text{partial}}$. The two types of integrals yield,%
\begin{equation}
I_{\text{full}}=\int_{\xi_{v_{k}}}^{\xi_{v_{k+1}}}S_{v}\left(  t\text{, }%
f_{v}\text{, }\lambda_{v}\text{, }\xi_{v}\text{, }E_{v}\right)  dt\text{,}%
\label{a1}%
\end{equation}
with $k=1$,$\cdots$, $E_{v}-1$, and
\begin{equation}
I_{\text{partial}}=\int_{\xi_{v_{k}}}^{t_{i}}S_{v}\left(  t\text{, }%
f_{v}\text{, }\lambda_{v}\text{, }\xi_{v}\text{, }E_{v}\right)  dt\label{a2}%
\end{equation}
with $k=1$,$\cdots$, $E_{v}-1$,$\,\ i=1$,$\cdots$, $n$,  and $\xi_{v_{k}}\leq
t_{i}<\xi_{v_{k+1}}$, respectively. The solutions to these two cases
presented in\ Eq. (\ref{a1}) and Eq. (\ref{a2}) can be found analytically and
result in the following expressions,
\begin{align}
I_{\text{partial}} &  =\frac{f_{v_{i+1}}}{2h_{v_{i}}}(t_{i}-\xi_{v_{i}}%
)^{2}+\frac{f_{v_{i}}(t_{i}-\xi_{v_{i}})}{2h_{v_{i}}}(2\xi_{v_{i+1}}-t_{i}%
-\xi_{v_{i}})\label{I_partial_sol}\\
&  -\frac{M_{v_{i}}(t_{i}-\xi_{v_{i}})}{2\lambda_{v_{i}}^{2}h_{v_{i}}}%
(2\xi_{v_{i+1}}-t_{i}-\xi_{v_{i}})\nonumber\\
&  +\frac{M_{v_{i}}}{\lambda_{v_{i}}^{3}\sinh(\lambda_{v_{i}}h_{v_{i}}%
)}\left\{  \cosh(\lambda_{v_{i}}h_{v_{i}})-\cosh\left[  \lambda_{v_{i}}%
(t_{i}-\xi_{v_{i+1}})\right]  \right\}  \nonumber\\
&  +\frac{M_{v_{i+1}}}{\lambda_{v_{i}}^{3}\sinh(\lambda_{v_{i}}h_{v_{i}}%
)}\left\{  \cosh\left[  \lambda_{v_{i}}(t_{i}-\xi_{v_{i}})\right]  -1\right\}
-\frac{M_{v_{i+1}}}{2\lambda_{v_{i}}^{2}h_{v_{i}}}(t_{i}-\xi_{v_{i}}%
)^{2}\text{,}%
\end{align}
and,
\begin{equation}
I_{\text{full}}=\frac{f_{v_{i}}h_{v_{i}}}{2}+\frac{f_{v_{i+1}}h_{v_{i}}}%
{2}-\frac{M_{v_{i}}h_{v_{i}}}{2\lambda_{v_{i}}^{2}}-\frac{M_{v_{i+1}}h_{v_{i}%
}}{2\lambda_{v_{i}}^{2}}+\frac{(M_{v_{i+1}}+M_{v_{i}})}{\lambda_{v_{i}}^{3}%
}\tanh\left(  \frac{\lambda_{v_{i}}h_{v_{i}}}{2}\right)  \text{.}%
\label{I_full_sol1}%
\end{equation}

\section{} \label{appendix II}
In this Appendix, being in the framework of Markov chain Monte Carlo
(MCMC) methods, we present some details on the Delayed Rejection Adaptive
Metropolis (DRAM) algorithm used to compute posterior distributions in our
work. DRAM is a hybrid algorithm where the concepts of Delayed Rejection
(DR) and Adaptive Metropolis (AM) are combined for enhancing the performance of the Metropolis-Hastings (MH) type MCMC algorithms \cite{haario2006dram}. In the basic MH algorithm, upon the rejection of
the new candidate at a (j + 1)th state, the sample remains at the current point
(j). However, when the Delayed Rejection is introduced, upon rejection, another new candidate z is drawn. Then, one finds out whether it can be accepted.
The AM feature, instead, is a global adaptive strategy which is combined with
the partial local adaptation caused by the DR strategy. The idea of AM is to
tune the proposal distribution at each time step depending on the past. In other words, the covariance matrix of the proposal distribution depends on the history
of the sample chain (sequence of samples). Usually the adaptation starts after
an initial period of time (0, t).

\begin{algorithm}[H]
	\caption{DRAM}
	\label{alg:DRAM}
	\begin{algorithmic}[1] 
		\FOR{$j=1:N$}
		\STATE Generate $y$ from $q\left(\Theta^{(j)},.\right)$ and $u$ from $U(0,1)$ \\
		\IF{$y$ is outside $\left(L,U\right)$}
		\STATE Reject $y$ (level 1)	
		\ELSE 
		\STATE 	Calculate $\alpha_1(\Theta^{(j)},y)$
		\IF{$u\leq \alpha_1(\Theta^{(j)},y)$}
		\STATE $\Theta^{(j+1)} = y$
		\ELSE
		\STATE Reject $y$ (level 2)
		\COMMENT{starts delayed rejection}	
		\STATE Generate new candidate $z$ from $q\left(y,.\right)$
		\IF{$z$ is outside $\left(L,U\right)$}
		\STATE Reject $z$ (level 3)
		\ELSE
		\STATE Calculate $\alpha_2(\Theta^{(j)},y,z)$ and $u$ from $U(0,1)$
		\IF{$u \leq\alpha_2(\Theta^{(j)},y,z)$}
		\STATE $\Theta^{(j+1)} = z$	
		\ELSE
		\STATE Reject $z$ (level 4)
		\STATE $\Theta^{(j+1)} = \Theta^{(j)}$
		\ENDIF
		\ENDIF
		\ENDIF
		\ENDIF
		\ENDFOR
		\STATE Return the values $\{\Theta^{\left(0\right)},\Theta^{\left(1\right)},\Theta^{\left(2\right)},\cdots,\Theta^{\left(N\right)}\}$.
	\end{algorithmic}	
\end{algorithm}

The basic steps of the DRAM algorithm is given in Algorithm \ref{alg:DRAM} with an arbitrary initial point $\Theta^{(0)}$. The acceptance probabilities at steps 6 and 15 are shown in Eqs.~\eqref{alpha1 dram} and~\eqref{alpha 2 dram} respectively.
\begin{subequations}
	\begin{align}
	\alpha_1\left(x,y\right) &= \begin{cases}
	\mbox{min}\left[\frac{\pi\left(y\right)q\left(y,x\right)}{\pi\left(x\right)q\left(x,y\right)},1\right] \quad &\mbox{if  } \pi\left(x\right)q\left(x,y\right)>0, \label{alpha1 dram}\\
	1 \quad &\mbox{otherwise}.
	\end{cases}\\
	\alpha_2\left(x,y,z\right) &= \begin{cases}
	&\mbox{min}\left[\frac{\pi\left(z\right)}{\pi\left(x\right)}\frac{q\left(z,y\right)}{q\left(x,y\right)}\frac{q\left(z,y,x\right)}{q\left(x,y,z\right)}\frac{[1-\alpha_1\left(z,y\right)]}{[1-\alpha_1\left(x,y\right)]},1\right]\\
	&\mbox{if } \pi\left(x\right)q\left(x,y\right)q\left(x,y,z\right)[1-\alpha_1\left(x,y\right)]>0\label{alpha 2 dram},\\
	1&\quad \mbox{otherwise}.
	\end{cases}
	\end{align}
\end{subequations}
$q(\theta)$ represents the proposal distribution and $\Theta^{(j)}$ represents the set of parameters of the proposal distribution at j-th iteration.

\end{document}